\documentclass[11pt]{article}
\usepackage[utf8]{inputenc}
\usepackage[OT1]{fontenc}

\usepackage[english]{babel}
\usepackage[backend=bibtex,natbib=true,citestyle = authoryear,style = authoryear,maxbibnames=10,maxcitenames=2,dashed=false]{biblatex}
\usepackage{hyperref}
\usepackage{xcolor}
\usepackage{amsmath,amsfonts,amssymb,amsthm}
\theoremstyle{remark}
\newtheorem{rmk}{Remark}

\usepackage{csquotes}
\usepackage{stmaryrd}
\usepackage[page]{appendix}
\usepackage{ifthen}
\usepackage{xargs}
\usepackage{tabularx}
\usepackage{multirow}
\usepackage{graphicx}

\usepackage{fullpage}
\usepackage{setspace}
\usepackage[ruled,vlined]{algorithm2e}
\usepackage{booktabs}

\usepackage[affil-it]{authblk}

\usepackage{pgfplots}
\usetikzlibrary{matrix}
\usepgfplotslibrary{groupplots}
\DeclareUnicodeCharacter{2212}{−}
\usepgfplotslibrary{groupplots,dateplot}
\usetikzlibrary{patterns,shapes.arrows}
\pgfplotsset{compat=1.5}
\usetikzlibrary{external}
\DeclareMathOperator*{\argmax}{arg\,max}

\newcolumntype{Y}{>{\centering\arraybackslash}X}

\theoremstyle{definition}

\theoremstyle{remark}

\newcommand\diff{\mathrm{d}}
\newcommand{\intervalleEntier}[2][1]{\llbracket #1,#2 \rrbracket}
\newcommandx{\dpart}[3][1=1,3=x]{\ifthenelse{\equal{#1}{1}}{\frac{\partial#2}{\partial#3}}{\frac{\partial^{#1}#2}{\partial^{#1}#3}}}

\title{LOB modeling using Hawkes processes \\ with a state-dependent factor}
\author[1]{Emmanouil Sfendourakis\thanks{emmanouil.sfendourakis@centralesupelec.fr}}
\author[2]{Ioane Muni Toke\thanks{ioane.muni-toke@centralesupelec.fr}}
\affil[1,2]{Laboratoire Mathématiques et Informatique pour la Complexité et les Systèmes, CentraleSupélec, Université Paris-Saclay, France}
\date{\today}

\begin{document}

\maketitle

\begin{abstract}
A point process model for order flows in limit order books is proposed, in which the conditional intensity is the product of a Hawkes component and a state-dependent factor. In the LOB context, state observations may include the observed imbalance or the observed spread. Full technical details for the computationally-efficient estimation of such a process are provided, using either direct likelihood maximization or EM-type estimation. Applications include models for bid and ask market orders, or for upwards and downwards price movements. Empirical results on multiple stocks traded in Euronext Paris underline the benefits of state-dependent formulations for LOB modeling, e.g. in terms of goodness-of-fit to financial data.
\end{abstract}


\section{Introduction}

The limit order book (LOB) of a financial asset is the structure that keeps track of the buy and sell intentions of the said asset. Modelling the flow of orders sent to an order book is an important challenge in the field of financial microstructure. Outputs of such models could be of interest to e.g., assess  high-frequency trading strategies, or optimal execution strategies. A description of trading mechanisms and order book dynamics can be found in e.g., \textcite{gould2013limit} or \textcite{abergel2016limit}, alongside some empirical observations and a review of some models.

Order flows are random events in time, and thus naturally modelled with point processes. Poisson processes can be simple candidates for LOB modelling \parencite{cont2010stochastic}, although they fail to grasp the complex dependencies that can be found in order flows \parencite{abergel2016limit}. Recently, Hawkes processes have become a popular class of point processes to describe the dynamics of the orders arrivals, because of their ability to catch the self- and cross-exciting effects of several types of orders empirically observed on financial markets. A review of the literature on the use of these processes in financial microstructure can be found in \textcite{bacry2015hawkes}. Let us just recall here that multivariate Hawkes processes are intensity-driven processes, and that their intensity depends on a kernel matrix $\phi(t)$, $t\in\mathbb R_+$, as follows: when an event of type $j$ occurs at time $s$, then a term  $\phi_{ij}(t-s)$ is added to the intensity of the events of type $i$ at time $t>s$. In standard Hawkes processes, the term added is non-negative (excitation effect) and exponentially decreasing (i.e. typically $\phi_{ij}(t-s) = \alpha_{ij} e^{-\beta_{ij}(t-s)}$). One reason for the popularity of this kernel is that it allows recursive calculations of the likelihood function, which is crucial in keeping a reasonable computational time for maximum likelihood estimations. Exponential decay is a strong modelling assumption, especially in the field of financial microstructure, where events may often exhibit multiple timescales, or long memory. Such effects can be partially grasped using sums of exponential terms in the kernel. Such sums may approximate power decays \parencite[see e.g.,][]{hardiman2013critical}, and \textcite{lallouache2016limits} shows that using more than one exponential term greatly improves statistical fits of Hawkes processes to market data. Positivity of the effect may also be a strong assumption, as inhibiting effects can also be found in LOB dynamics. Non-linear Hawkes processes must be developped to take such effects into account \parencite{lu2018high}.

A main drawback of these Hawkes-based models is that their intensity does not depend on the current state of the limit order book, although empirical observations suggests that the state of the LOB influences the order flows. \textcite{HuangQueueReactiveModel} show that the rate of arrival of limit orders depend on the size of the queue. \textcite{munitoke2017modelling} present a parametric model for intensities depending only on the state of the order book and show that the rate of arrival of market orders is much higher when the spread is low. However, these models fail to catch excitation and clustering of the order flow. An attempt at defining state-dependent Hawkes processes for LOB modelling can be found in \textcite{morariu2018state}, where the intensity is computed with a Hawkes kernel varying with the state of the order book. Another related model is proposed in \textcite{wu2019queue}, which extends \textcite{HuangQueueReactiveModel} with Hawkes models depending on the size of the queues.

In this paper, we present an alternative state-dependent Hawkes model, inspired by the ratio statistics of \textcite{munitoke2020analyzing}, and where the intensity of the process is the intensity of a classical Hawkes process multiplied by an exponential factor depending on the observed state. Section \ref{sec:msdHawkes} presents this state-dependent Hawkes process with exponential state factor, which we call msdHawkes throughout the paper. We provide two estimation procedures for the msdHawkes processes, namely a direct maximum-likelihood estimation and an expectation-maximization algorithm. We compare the msdHawkes formulation with other state-dependent Hawkes-based LOB models and find it to be quite flexible in its ability to incorporate state covariates and multiple exponential kernels. Section \ref{sec:applications} provides empirical results. We describe the application of the msdHawkes model to different kind of two-dimensional LOB order flows: bid and ask market orders, and also upwards and downwards price movements. In both cases, we analyze the goodness-of-fit of state-dependent Hawkes models to a sample of 36 stocks traded on Euronext Paris during the year 2015. Furthermore, the msdHawkes processes are analyzed via the empirical estimates of the parameters, the endogeneity measure, as well as an out-of-sample prediction exercise. All technical details and computations needed to efficiently implement msdHawkes models are provided in Appendices \ref{appendix:MLE} to \ref{appendix:miniExample}.

\section{Hawkes process with state-dependent factor (msdHawkes)}
\label{sec:msdHawkes}

\subsection{Model definition}
Let $(\Omega, \mathcal F, (\mathcal{F}_t)_{t \geq 0}, \mathbf P)$ be a filtered probability space on which our processes will be defined.
Let $d_x \in \mathbb{N}$. Let $(X_t)_{t \geqslant 0}$ be an adapted, piecewise constant, left-continuous stochastic process, with values in $[-1, 1]^{d_x}$. $(X_t)_{t \geqslant 0}$ denotes the observable state space. We may remark that the piecewise constant hypothesis is consistent with financial microstructure framework: the LOB state changes as the result of order submission and remains constant between these events. State covariates considered in this paper, such as the spread or the imbalance, are indeed piecewise constant processes. 
Let $d_e\in \mathbb{N}^*$. Let $(N_t)_{t \geqslant 0}$ be a $d_e$-dimensional counting process with stochastic intensity $\lambda$. Recall that if $N$ were a standard Hawkes process, then we would have $\lambda=\lambda^H$ with
\begin{equation}
    \label{eq:standardHawkes}
    \forall t \geqslant 0, \quad \lambda^H(t) = \nu + \int_{]0, t[} \phi(t-s) \cdot \diff N_s,
\end{equation}
where $\nu \in \left(\mathbb{R}_+\right)^{d_e}$ is the baseline intensity and $\phi = (\phi_{ee'})_{e,e'=1,\ldots,d_e}$ the $d_e \times d_e$ matrix kernel function, assumed non-negative and locally integrable.

Let us now define our state-dependent formulation of the Hawkes process. A Hawkes process with (multiplicative) state-dependent factor (from now on abbreviated msdHawkes process) is defined as a point process $N$ with an intensity $\lambda=(\lambda^1,\ldots,\lambda^{d_e})$ of the form:
\begin{equation}
    \label{eq:msdHawkes}
    \forall e \in \intervalleEntier{d_e}, \quad \forall t \geqslant 0, \quad \lambda^e(t) = \lambda^{H, e}(t)\exp\left(\left< \theta^e, X_t \right> \right),
\end{equation}
where $\lambda^{H,e}$ is the $e$-th coordinate of a standard Hawkes intensity given by \eqref{eq:standardHawkes}, and $(\theta^1,\ldots,\theta^{d_e}) \in (\mathbb{R}^{d_x})^{d_e}$ are the state coefficients.
Throughout this paper, we will use multiple exponential kernels defined as:
\begin{equation}
    \label{eq:kernel}
    \forall e, e' \in \intervalleEntier{d_e}, \quad \forall t \geqslant 0, \quad \phi_{ee'}(t) = \sum_{n=1}^{d_n} \alpha_{ee'}^n e^{-\beta_{ee'}^n t},
\end{equation}
where $d_n \in \mathbb{N}^*$ and $\alpha_{ee'}^n, \beta_{ee'}^n > 0$, which gives a full parametrization of the model. As recalled in the introduction, these kernels can model effects with multiple timescales, while still allowing for recursive computations of the log-likelihood.

In what follows, we will try to understand the msdHawkes model via numerical simulations and applications to empirical data. All these numerical results are strong hints that the simulation and estimation methods are sound and reliable in this context. Theoretical results regarding existence, uniqueness and stability of such point processes, although desirable, are not trivial and tackling these issues is beyond the scope of this paper. Coupling between the counting processes and the state space is not explicitly studied here. Such an issue is discussed in \citet{morariu2018state} in the context of another state-dependent Hawkes model, which we introduce below in Section \ref{subsec:ksdHawkes}. As for stability, observe that if the state factor $\exp\left(\left< \theta^e, X_t \right> \right)$ is bounded and its supremum sufficiently small, then the msdHawkes intensity is dominated by the intensity of a sub-critical standard Hawkes process. With this assumption, a stationary state-process should be sufficient to obtain a non-explosive stationary msdHawkes process. Such an ideal case may however not be met in practice, as we will see below in Section \ref{subsec:endogeneity}: the endogeneity ratio in empirical experiments temporarily visits supercritical regions. In such cases, one may conjecture that the process is stable if it spends sufficient time in, or jumps sufficiently often into, a sub-critical state. Numerical simulations on very long horizons with parameters fitted on empirical data are comforting with this respect, but to our knowledge, these mathematical issues are yet to be resolved. The usefulness of the msdHawkes model we demonstrate in the following sections is hopefully an incentive for further mathematical studies.

\subsection{Maximum likelihood estimation}

msdHawkes processes can be estimated by maximizing the likelihood of an observation of the process in an interval $[0, T]$. The general form of the log-likelihood of a point process parametrized by $\vartheta$ with intensity $\lambda_{\vartheta}$ is \parencite{rubin1972regular}:
\begin{equation}
    \label{eq:pointProcessLogLik} L_T(\vartheta) = \sum_{e=1}^{d_e} \left[-\int_0^T  \lambda^e_{\vartheta} (s) \diff s  + \int_{]0,T]} \log\left( \lambda_{\vartheta}^e (s) \right) \diff N^e_s \right].
\end{equation}
The log-likelihood of a msdHawkes process with intensity \eqref{eq:msdHawkes} and kernel \eqref{eq:kernel} is thus written:
\begin{equation}
    \begin{split}
     L_T(\nu, \alpha, \beta, \theta) = \sum_{e=1}^{d_e} \bigg[ - &\int_0^T \lambda^{H,e} (s) e^{\left< X(s), \theta^e \right>}\diff s + \int_0^T \log(\lambda^{H,e} (s))\diff N^e(s)\\
     & + \int_0^T \left< X(s), \theta^e \right> \diff N^e(s)\bigg].
\end{split}
\label{eq:msdloglik}
\end{equation}
Several remarks can be made regarding this log-likelihood.
Firstly, observe that each term $e$ of the sum \eqref{eq:msdloglik} depends only on the set of parameters
\begin{equation}
    \lbrace \nu_e \rbrace \cup \lbrace \alpha_{ee'}^n: e' \in \intervalleEntier{d_e}, n \in \intervalleEntier{d_n}\rbrace \cup \lbrace \beta_{ee'}^n: e' \in \intervalleEntier{d_e}, n \in \intervalleEntier{d_n}\rbrace \cup \lbrace \theta^e \rbrace.
\end{equation}
These sets being disjoint, the maximization can therefore be done separately for each term of the sum.
Secondly, note that when $d_n > 1$, exponential terms in the kernel \eqref{eq:kernel} can be swapped and $\phi$ would still be the same, making the model not identifiable. To correct that, we force $\beta_{ee'}^{1} > \dots > \beta_{ee'}^{d_n}$. Thus, the first terms in \eqref{eq:kernel} account for a short-term excitation and the last ones for the long-term excitation effects.
Finally, the exponential form of the kernel allows the log-likelihood to be computed recursively, one function evaluation being computed in $O(k)$ operations instead of $O(k^2)$, where $k$ is the total number of jumps. Numerical maximization can be carried out with gradient-based methods, and recursive computations are also available for gradients. All computations and detailed results are given in Appendix \ref{appendix:LogLik} for the log-likelihood and Appendix \ref{appendix:gradLogLik} for the gradients. Some numerical examples are given below.

\subsection{Expectation-Maximization likelihood estimation}

Direct maximization of the log-likelihood of multi-dimensional Hawkes processes may be numerically challenging \parencite{lu2018high}. Expectation-Maximization (EM) algorithms have been developed for standard Hawkes processes, by taking advantage of the branching structure to develop a complete likelihood maximization problem \parencite{veen2008estimation}. In this section we develop a EM-type algorithm for the estimation of msdHawkes processes with exponential kernels \eqref{eq:standardHawkes}-\eqref{eq:kernel}.
We consider the branching structure with the lowest granularity by isolating each exponential of each kernel. This decomposition of events is similar to marking the events, or to extending the type of the events from $e=1,\ldots,d_e$ to $(e,n)$, $e=1,\ldots,d_e$, $n=1,\ldots,d_n$. For the $i$-th event of type $e$, we set
\begin{equation}
u^{e}_i = \begin{cases}
	(e,i) & \text{if it is an exogenous event},
	\\
	(e',n',j) & \text{if its immediate ancestor is the $j$-th event of type $(e',n')$}.
\end{cases}
\end{equation}
For any $(e,i)$ and $(e',j)$ such that $t^{e'}_j<t^{e}_i$,
\begin{equation}
\begin{cases}
	\mathbf P\left(u^{e}_i=(e,i)\mid \mathcal F_{t^e_i}\right) = \frac{\nu_e e^{\langle \theta^e,X_{t^e_i-} \rangle}}{\lambda^e(t^e_i-)} = \frac{\nu_e}{\lambda^{H,e}(t^e_i-)},
	\\
	\mathbf P\left(u^{e}_i=(e',n',j)\mid \mathcal F_{t^e_i}\right) = \frac{e^{\langle \theta^e,X_{t^e_i-} \rangle}}{\lambda^e(t^e_i-)} \alpha_{ee'}^{n'} e^{-\beta_{ee'}^{n'}(t^e_i-t^{e'}_j)}
	= \frac{\alpha_{ee'}^{n'} e^{-\beta_{ee'}^{n'} (t^e_i-t^{e'}_j)}}{\lambda^{H,e}(t^e_i-)},
\end{cases}
\end{equation}
One can thus define for a msdHawkes sample with horizon $T$ a complete log-likelihood $L^c_T(\nu,\alpha,\beta,\theta\mid u) = \sum_{e=1}^{d_e} L^{c,e}_T(\nu,\alpha,\beta,\theta\mid u)$, where for any $e=1,\ldots,d_e$:
\begin{align}
L^{c,e}_T(\nu,\alpha,\beta,\theta\mid u)
= &
-\int_0^T \nu_e e^{\langle \theta^e,X_{t} \rangle}\,\diff t
+ \sum_{t^{e}_i} \mathbf 1_{u^{e}_i=(e,i)} \log \nu_e + \sum_{t^{e}_i} \mathbf 1_{u^{e}_i=(e,i)} \langle \theta^e,X_{t^e_i-} \rangle  
\nonumber \\ & 
+ \sum_{e'=1}^{d_e} \sum_{t^{e'}_j} \sum_{n'=1}^{d_n} \Bigg[
-\int_{t^{e'}_j}^{T} \alpha_{ee'}^{n'} e^{-\beta_{ee'}^{n'}(t-t^{e'}_j)} e^{\langle \theta^e,X_{t} \rangle}\,\diff t
\nonumber \\ & 
 + \sum_{t^{e}_i : t^{e'}_j<t^{e}_i} \mathbf 1_{u^{e}_i=(e',n',j)} \langle \theta^e,X_{t^{e}_i-} \rangle
\nonumber \\ & 
 + \sum_{t^{e}_i : t^{e'}_j<t^{e}_i} \mathbf 1_{u^{e}_i=(e',n',j)} \log\left(\alpha_{ee'}^{n'} e^{-\beta_{ee'}^{n'} (t^{e}_i-t^{e'}_j)}  \right)
\Bigg].
\end{align}
Taking the expectation and computating the partial derivatives with respect to the parameters yields update equations for an EM-type algorithm. Full EM-algorithm for msdHawkes processes \eqref{eq:standardHawkes}-\eqref{eq:kernel} is given in Appendix \ref{appendix:EM_algo}. 

\subsection{Numerical illustrations}
\label{subsec:numerical}

In this section we provide illustrations of the estimation algorithms on simulated data. On all examples, simulated data is obtained via the standard thinning method for point processes. 
In a first experiment, we check the performances of the methods described above on a msdHawkes model with $d_e=2$, $d_n=1$ and $d_x=2$. In this experiment and the following ones, the state process $X$ takes independent values in $[-1,1]^2$ and jumps randomly following an homogeneous Poisson process with intensity $1$. Table \ref{table:EstimationExamples} presents estimation results for direct log-likelihood maximization with two gradient-based optimization methods (L-BFGS-B or TNC) and for the EM algorithm.
\begin{table}[htbp]
\small
\begin{center}
\begin{tabular}{|c|c|c|c|c|c|c|c|}
\hline
 & $\hat\nu_{1}$ & $\hat\alpha_{11}^{1}$ & $\hat\alpha_{12}^{1}$ & $\hat\beta_{11}^{1}$ & $\hat\beta_{12}^{1}$ & $\hat\theta^1_{1}$ & $\hat\theta^1_{2}$
\\
True value & 0.500 & 4.000 & 0.400 & 8.000 & 2.000 & 0.250 & -0.250 \\ \hline
L-BFGS-B & 0.500 & 3.992 & 0.408 & 8.037 & 1.935 & 0.242 & -0.247 
\\
 & \textit{(0.045)} & \textit{(0.372)} & \textit{(0.196)} & \textit{(0.642)} & \textit{(1.316)} & \textit{(0.054)} & \textit{(0.072)}
\\ \hline
TNC & 0.500 & 3.938 & 0.401 & 7.980 & 2.051 & 0.256 & -0.245 
\\
 & \textit{(0.048)} & \textit{(0.436)} & \textit{(0.195)} & \textit{(0.749)} & \textit{(1.378)} & \textit{(0.055)} & \textit{(0.065)} 
\\ \hline
EM & 0.505 & 3.983 & 0.404 & 7.911 & 2.081 & 0.249 & -0.246
\\
 & \textit{(0.034)} & \textit{(0.397)} & \textit{(0.266)} & \textit{(0.705)} & \textit{(1.424)} & \textit{(0.062)} & \textit{(0.067)}
\\ \hline\hline
 & $\hat\nu_{2}$ & $\hat\alpha_{21}^{1}$ & $\hat\alpha_{22}^{1}$ & $\hat\beta_{21}^{1}$ & $\hat\beta_{22}^{1}$ & $\hat\theta^2_{1}$ & $\hat\theta^2_{2}$ 
\\
True value & 0.250 & 1.000 & 0.200 & 8.000 & 2.000 & -0.250 & 0.250 \\ \hline
L-BFGS-B & 0.254 & 0.991 & 0.212 & 8.011 & 2.352 & -0.245 & 0.245
\\
 & \textit{(0.030)} & \textit{(0.239)} & \textit{(0.154)} & \textit{(1.859)} & \textit{(1.878)} & \textit{(0.117)} & \textit{(0.111)} 
\\ \hline
TNC & 0.247 & 1.024 & 0.217 & 8.216 & 2.229 & -0.253 & 0.252 
\\
 & \textit{(0.034)} & \textit{(0.235)} & \textit{(0.132)} & \textit{(1.914)} & \textit{(1.395)} & \textit{(0.117)} & \textit{(0.110)}
\\ \hline
EM & 0.249 & 1.022 & 0.216 & 8.045 & 2.065 & -0.246 & 0.251 
\\
 & \textit{(0.027)} & \textit{(0.215)} & \textit{(0.141)} & \textit{(1.679)} & \textit{(1.829)} & \textit{(0.112)} & \textit{(0.106)} \\ \hline
\end{tabular}
\end{center}
\caption{Estimation results for various methods on simulated msdHawkes processes. See text for details. }
\label{table:EstimationExamples}
\end{table}
For each parameter, reported values are median and interquartile distance (in parentheses) for 120 simulated samples with given true values and horizon $T=1000$. For each sample, optimization starts from a random point. The three methods gives comparable results, and strongly outperform basic Nelder-Mead log-likelihood maximization (not reported because of these poorer performances). 

\begin{figure}
\centering
\includegraphics[width=\textwidth]{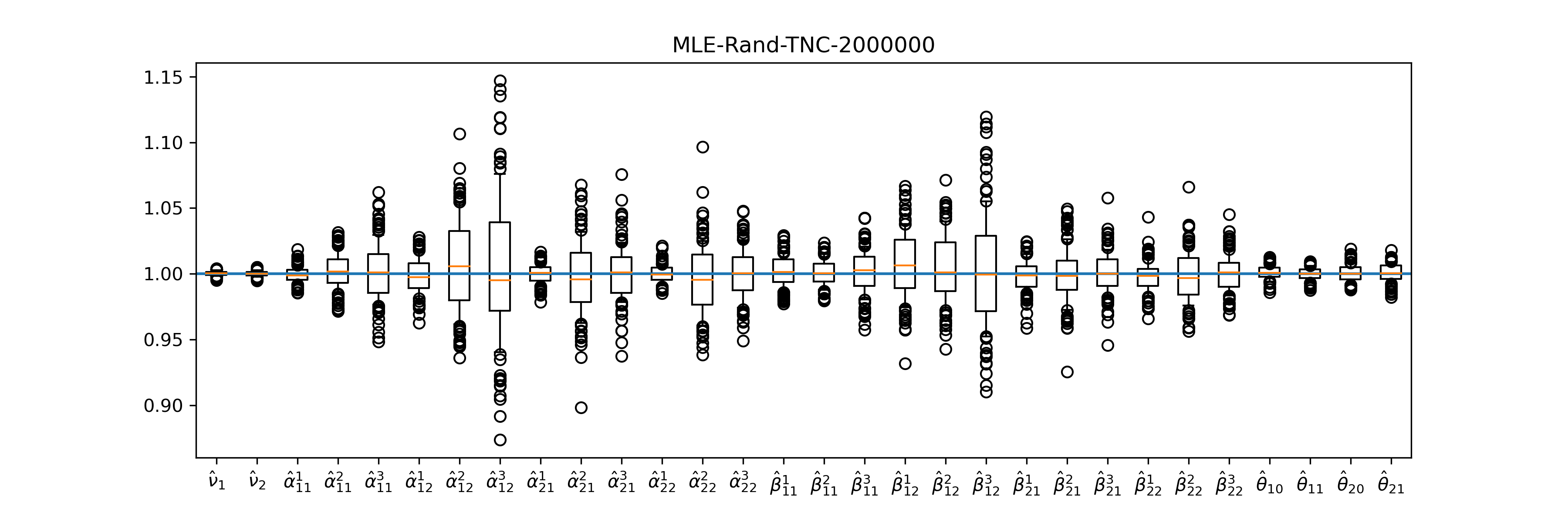}
\caption{Scaled distribution of estimated values for a msdHawkes model on simulated data with $d_e=2$, $d_x=2$ and $d_n=3$ (30 parameters). See text for details.}
\label{fig:EstimationExample}
\end{figure}

In a second experiment, we provide an example with multiple exponential kernels. We simulate for various horizon values 120 samples of a msdHawkes model with $d_e=2$, $d_x=2$ and $d_n=3$. We then estimate the model using direct likelihood maximization with a TNC maximization routine, which appeared effective in the previous experiment. Figure \ref{fig:EstimationExample} plots the full boxplots (scaled by the true value of the parameter) of the 120 estimated values for each of the 30 parameters for $T=2\times 10^6$. Baseline intensities and state parameters are very efficiently estimated. Hawkes estimates exhibit a larger variance, but all median values are within less than $1\%$ of the true value.
As a complement, Figure \ref{fig:StdDeviationConvergence} in Appendix \ref{appendix:ComplNum} checks the decreasing of the standard deviation of the 30 estimates in $\mathcal O(T^{-1/2})$.

Direct log-likelihood maximization has been used in the second experiment because of computational performance.
Indeed, gradient-based direct log-likelihood maximization clearly outperforms the EM estimation in terms of computational times: in the first experiment where a typical sample has roughly $10^3$ points, L-BFGS-B and TNC methods take less than 1 second, while EM method needs on the same hardware in average 4500 to 11000 seconds depending on the tolerance for the stopping criterion. It has been shown on (different) Hawkes-like processes that EM algorithm may help get more robust results, and avoid local maxima of the log-likelihood, when one does not know anything about the initial parameters \citep{mark2020robust}. In this sense, this first experiment is rather positive regarding the performances of the direct likelihood maximization. Direct maximization with gradient-based methods will thus be used for the empirical results presented in Section \ref{sec:applications}. In order to lessen the possibility of finding a local maximum of the non-convex likelihood, grid searches may be used and genetic algorithms have also been suggested in this context \citep{lu2018high}. These methods are not used in our numerical experiments ; we nonetheless obtain good results in these experiments without such improvements, even with multiple exponential kernels. In the context of empirical data in Section \ref{sec:applications}, we use for each sample up to 12 optimizations with different random starting points and keep the best results in terms of likelihood.

\begin{figure}
\centering
\includegraphics[width=0.6\textwidth]{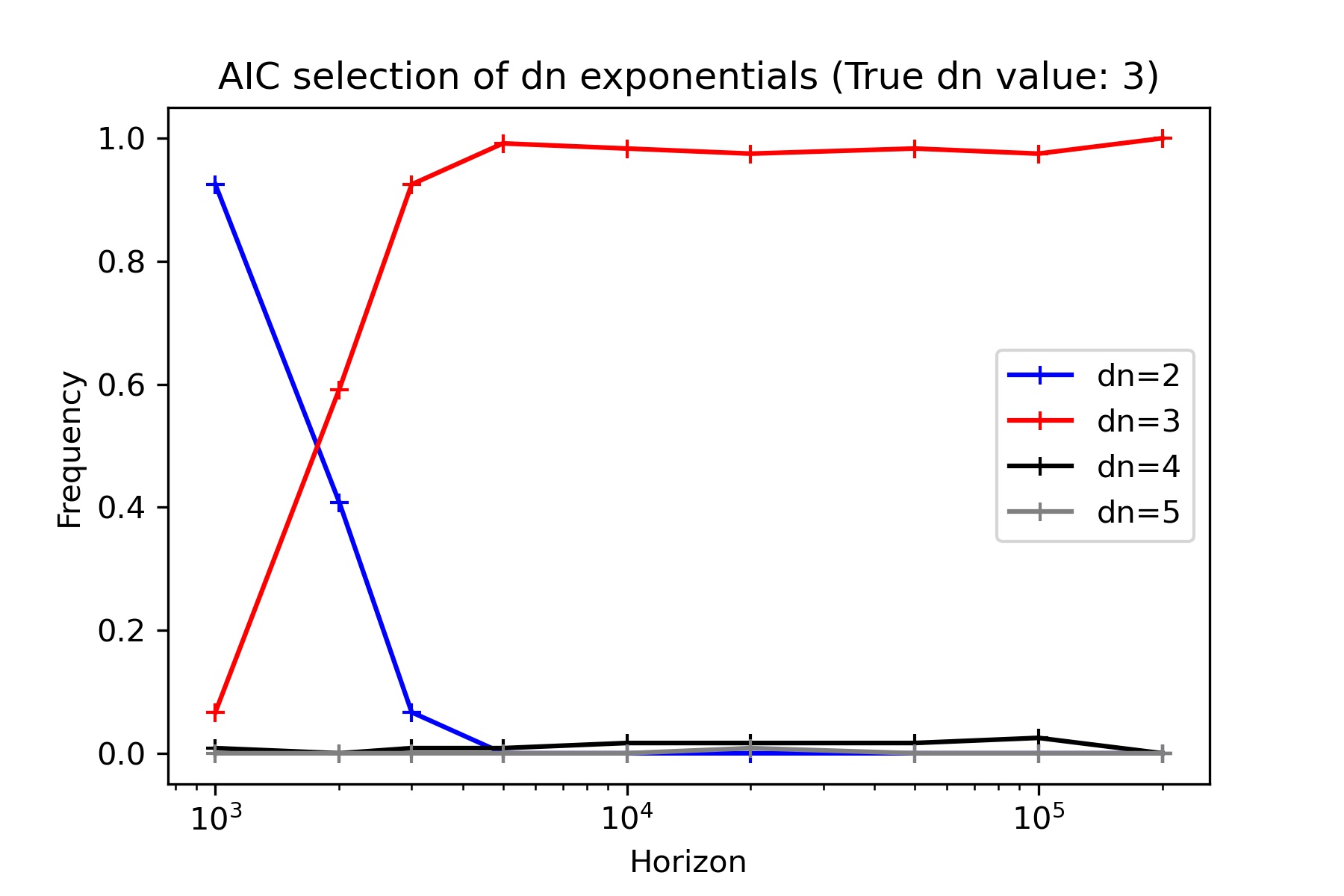}
\caption{Frequency of selection of a msdHawkes model with $d_n$ exponentials on data simulated with $d_n=3$. See text for details.}
\label{fig:MultipleExp-AICSelection}
\end{figure}

In a third experiment, we verify the ability to estimate the correct number of exponential terms in a kernel via AIC. Using the same numerical values as in the previous experiment, we simulate for various horizon values 120 samples of a msdHawkes model with $d_e=2$, $d_x=2$ and $d_n=3$. We then estimate on each sample a msdHawkes model with $d_e=2$, $d_x=2$ and $d_n\in\{1\ldots,5\}$. Figure \ref{fig:MultipleExp-AICSelection} plots the frequencies of selection of each model using the AIC criterion.
In short samples with few points, a smaller number of exponentials is preferred, then AIC quickly catches the correct number of exponentials. For $T\geq 5000$, AIC selects the correct number of exponentials in nearly all samples. Note that in this experiment the longest characteristic time of the kernel is $(\min_{e,e',k} \beta_{ee'}^k)^{-1}=1$ unit of time.

\begin{figure}
\centering
\includegraphics[width=0.6\textwidth]{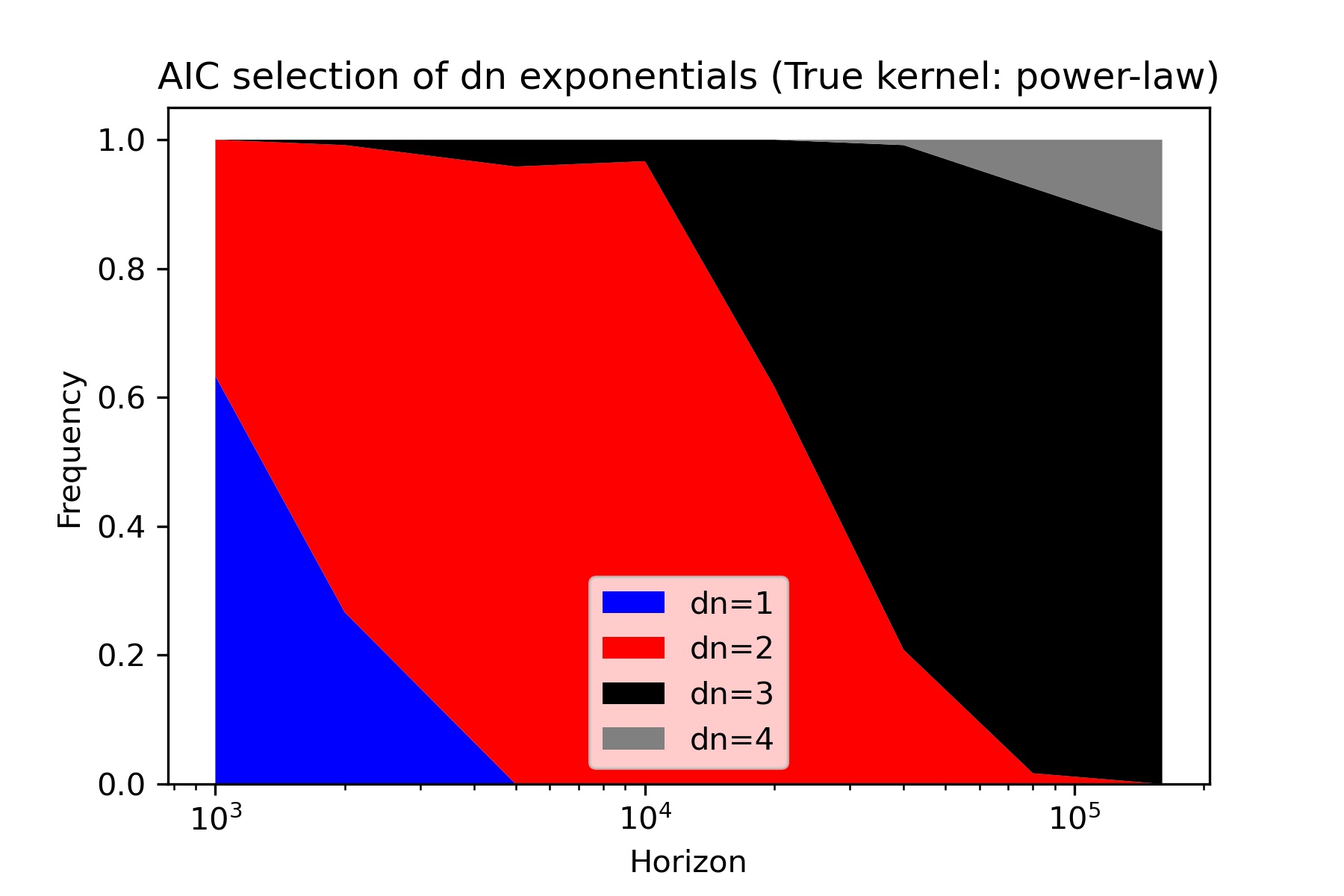}
\caption{Stacked frequencies of selection of a msdHawkes model with $d_n$ exponentials on data simulated with a power-law kernel. See text for details.}
\label{fig:PowerLaw-AICSelection}
\end{figure}

In a fourth numerical experiment, we use the same protocol but with a power-law kernel. In detail, we simulate 120 samples of a msdHawkes model $d_e=2$, $d_x=2$, and in which the kernels are of the form $\phi_{ee'}(t) = \alpha_{ee'}\left(1+t/\tau_{ee'}\right)^{-(1+\beta_{ee'})}$ instead of the sum of exponentials of Equation \eqref{eq:kernel} used so far. Numerical values used in this experiment are reported in Appendix \ref{appendix:ComplNum}. Note that such kernels do not satisfy useful recursive formula for computing intensities, so that these simulations are rapidly computationally expensive. Then for each simulated sample, we estimate a msdHawkes model with $d_e=2$, $d_x=2$ and $d_n\in\{1\ldots,5\}$. Figure \ref{fig:PowerLaw-AICSelection} plots the stacked frequencies of selection of each model using the AIC criterion with respect to the horizon.
As in the previous experiment, few exponential terms are selected on short simulations. Then as the horizon increases, the number of exponential terms preferred by AIC increases. In this experiment with $\beta_{ee}=1$ (self-excitation) and $\beta_{ee'}=2$ (cross-excitation), $d_n=1$ is the most selected up to $T=10^3$, then $d_n=2$ is the most selected for horizons roughly in the range $(2\times 10^3, 2\times 10^4)$, then $d_n=3$ is the most selected. $d_n=4$ begins to appear for $T\geq 10^5$ : it is selected roughly in $15\%$ of the samples for $T=160\,000$, i.e. approximately $520\,000$ events. Larger horizons are computationally too expensive to reach the range where we might expect $d_n=4$ to be the most selected. These observations are in accordance with the fact that the number of exponential terms needed to approximate a power-law increases with the length of the tail one wants to approximate \citep{bochud2007optimal}.

\subsection{Comparison with other state-dependent Hawkes formulations}
\label{subsec:ksdHawkes}

\textcite{morariu2018state} propose a different approach to define a state-dependent Hawkes model. They define a kernel-based state-dependent Hawkes process (from now on abbreviated ksdHawkes) as the couple $(N, Y)$, $N$ being a $d_e$-variate ($d_e \in \mathbb{N}^*$) point process and $X$ being the state function, with $Y$ being piecewise constant, right-continuous and valued in $\intervalleEntier{\delta_x}$ ($\delta_x \in \mathbb{N}^*$), and $N$ having the intensity $\lambda$:
\begin{equation}
    \forall t \geqslant 0, \quad \lambda(t) = \nu + \int_{]0,t[} \phi_{\cdot \cdot Y(s)} (t-s) \cdot \diff N_s,
\end{equation}
where $\nu \in \mathbb{N}^{d_e}$ is the baseline intensity and $\phi_{\cdot \cdot 1}, \dots, \phi_{\cdot \cdot \delta_x}$ are $d_e \times d_e$ kernel matrices (again assumed non-negative and locally integrable). In this setting, the jumps of $Y$ must occur at the same time as the jumps of $N$, in the following way: if the $e$-th coordinate of $N$ jumps at $t$, $Y$ jumps at $t$ according to a $\delta_x \times \delta_x$ transition matrix $\psi_{e \cdot \cdot}$. Kernels are of the form $\phi_{eex}(t) = \alpha_{ee'x}e^{-\beta_{ee'x}t}$ (single exponential). Estimation is done by maximizing the likelihood, and a Python library \texttt{mpoints} \parencite{mpoints} is available to carry out this task. Empirical results are presented for a two-class model ($d_e=2$) of level-1 orders (up and down price pressure) and the state space is either the spread ($\delta_x=2$) or a discretized imbalance ($\delta_x=5$).

\textcite{wu2019queue} include a model similar to the one proposed here, in which the intensity of submission of an order of type  $e \in \{1,\ldots,d_e\}$ in the LOB is a function of the size of the best bid and ask queues $q^B(t)$ and $q^A(t)$:
\begin{equation}
 \lambda_e(t) = f_e \left(q^B(t), q^A(t)\right) \left(\nu_e + \sum_{e'=1}^{d_e} \int_0^t \varphi_{ee'} (t - s) \, dN_{e'}(s) \right).
\end{equation}
This model is called queue-reactive Hawkes model by its authors (qrHawkes from now on). Empirical results are provided for a $8$-dimensional model of the  best limits with a discretized version of the state process $X(t)=(q^A(t)$, $q^B(t))$. 

Designs of these models are different. In the ksdHawkes formulation, the intensity depends on the history of the state process: the excitation brought by an order depends on the state of the order book at the time this order was submitted, not the current one. Knowledge of the state of the order book at a time $t$ is not enough to compute the intensity, while in the msdHawkes and qrHawkes models only the current state is relevant. One can  observe that the main difference between msdHawkes and qrHawkes models is that the latter explicitly assume that the state is provided by an arbitrary function of the queue sizes while the former explicitly assume an exponential dependence on an arbitrary state vector.

msdHawkes offers a flexible and parsimonious modeling of the state-representation. Indeed, LOB state is represented with $d_x$ variables, with either discrete or continuous values, while state space in the ksdHawkes or qrHawkes models must be discrete and finite, with $\delta_x$ total number of possible states. The cost of the kernel structure the ksdHawkes formulation is that the dimension of the parameter space grows linearly with the cardinal of the state space with coefficient $d_e^2$. More precisely, estimation is done in two separate steps: once the $d_e \,\delta_x^2 $ state transition parameters have been estimated, the total number of Hawkes parameters to be estimated is $d_e + 2 \,d_e^2 \,\delta_x$ parameters. The total number of different states $\delta_x$ must therefore be kept low to have a reasonable number of parameters. The qrHawkes formulation also requires a discretized and finite state-space in the non-parametric space factor. \citep{wu2019queue} expresses queue sizes ($q^A, q^B$) in quantiles, so that $\delta_x=25$, and spread is not included in the model. In the msdHawkes model, increasing the state space by one covariate, either discrete or continuous, only adds $d_e$ parameters, which makes the msdHawkes model quite parsimonious. 

For a similar reason of parsimony, the number of exponential terms in the kernels of the ksdHawkes model is kept equal to one, and a higher number of exponential terms is not tested (neither in \citet{morariu2018state} nor in our benchmarks using the original code), since each added exponential in the kernel increases the number of parameters by $2 \,d_e^2 \,\delta_x$ parameters, with $\delta_x$ the cardinal of the state space. In the msdHawkes model, each added exponential increases the number of parameters by $2\,d_e^2\, d_x$, $d_x$ being just the dimension of the state space ($d_x=2$ in the following applications). We will see in Section \ref{sec:applications} that the ability to easily add exponential terms in the kernels leads to significant differences in fitting results on empirical LOB data.

\section{Applications of the msdHawkes model to LOB data}
\label{sec:applications}

\subsection{msdHawkes models and benchmarks}

In this section, we use the msdHawkes model on limit order book data. Analysis is done on two different types of two-dimensional processes ($d_e=2$). In the first case, the point process $N$ counts all market orders, bid on the first coordinate, and ask on the second coordinate. This first case will be called `Market' case from now on. In the second case, the point process $N$ counts all orders that change the price (\emph{aggressive} orders), irrespective of their type (limit orders, market orders, cancellations). More precisely, first coordinates count upwards mid-price movements (ask market orders matching the full first limit, bid limit orders between the best quotes, ask cancellation orders of the full first limit), while the second coordinate count downwards mid-price movements (bid market orders matching the full first limit, ask limit orders between the best quotes, bid cancellation orders of the full first limit). This second case will be called `Aggressive' case from now on.

\begin{figure}
\centering
\includegraphics[width=0.7\textwidth]{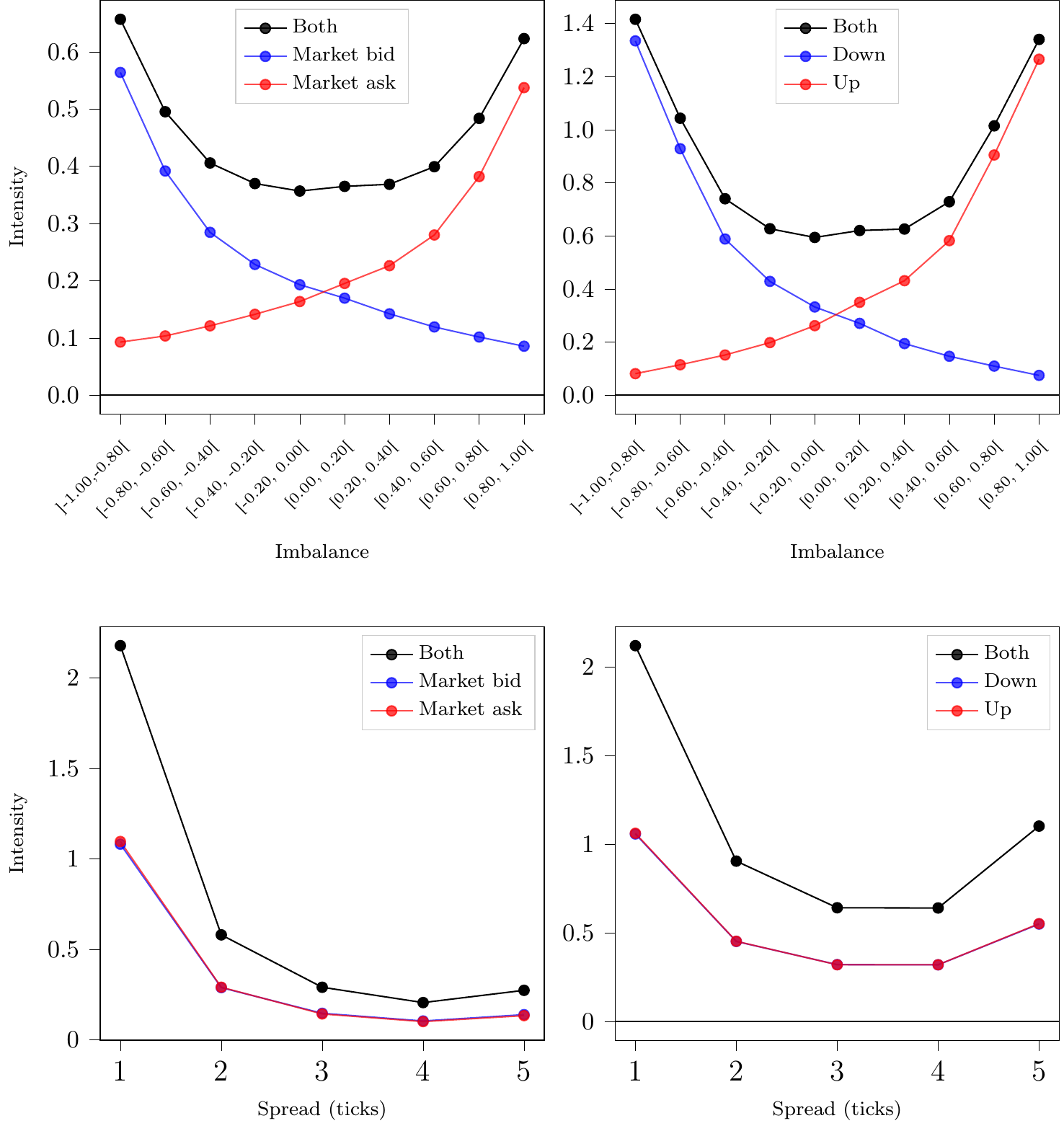}
\caption{Empirical intensities of `Market' orders (left column) and `Aggressive' orders (right column) as a function of the imbalance (top row) or the spread (bottom row) for the stock SOGN.PA traded in Paris in 2015.}
\label{fig:empirical_intensities}
\end{figure}

For each of these applications, we will represent the state space with functions of two main state variables: the bid-ask spread and the imbalance. Recall that if at time $t$, the best ask quote is at price $a(t)$ with aggregated size $q^A(t)$ and the best bid quote is at price $b(t)$ with aggregated size $q^B(t)$, then the spread is $s(t)=a(t)-b(t)$ and the imbalance is $i(t) = \frac{q^B(t) - q^A(t)}{q^B(t) + q^A(t)}$. These variables are known to influence the order flow in a limit order book \citep{munitoke2017modelling}. As a reminder, Figure \ref{fig:empirical_intensities} illustrates such dependencies by plotting empirical intensities in the `Market' and `Aggressive' cases as a function of the observed imbalance or spread at the time of submission. A large spread reduces the rate of transactions at market price, as a trader will prefer to submit a limit order between the best quotes, which is likely to be executed quickly, and at a better price. An imbalance close to one is more likely to lead to a price increase, as it is easier to consume the best ask queue in this case \citep{lipton2013trade}.
By definition, imbalance ($I$) takes values in $[-1,1]$. State-representation in msdHawkes models allows to include it in the model as is as one coordinate of the state space. As for the spread, three alternatives are tested in these experiments to include a spread covariate as a coordinate of the state space:
\begin{itemize}
    \item[-] S1: the covariate takes the value $-1$ if the spread in less or equal than the median spread, $1$ otherwise ;
    \item[-] S2: the covariate takes the value $-1$ if the spread is equal to one tick, $1$ otherwise ;
    \item[-] S3: let $p$ be the probability distribution function of the spread (in ticks), the covariate is $p$ linearly transformed (positively) so that its minimum is $-1$ and its maximum is 1 (if the spread is equal to $s$ ticks, the covariate is $\frac{s - \min p}{\max p - \min p}$).
\end{itemize}
Specification S2 should be useful for rather large tick stocks, where the spread is very often equal to one tick. Specification S3 generalizes this approach by defining regimes depending on the frequency of observation of a given spread value. Specifications S1 and S3 need the knowledge of the probability distribution of the spread. As a proxy it will be taken here in-sample (one year of data), assuming this distribution remains constant through time.
In the rest of the paper, we test several model specifications with varying number of exponentials in the Hawkes kernels. For easier readability, a short code is associated with each model. For example, 'msd-I-S3-2' denotes a msdHawkes model with imbalance and spread S3 as covariate, and $d_n=2$ exponentials in each Hawkes kernel. 'msd-I-S' denotes msdHawkes models with imbalance, an unspecified spread covariate, and an unspecified number of exponentials in the model. Precise meaning in this case (best model, average result, etc.) should be clear from context.

Finally, msdHawkes models will be compared to several benchmark models, namely standard Hawkes processes (coded 'std') and ksdHawkes models (coded 'ksd'). As discussed in Section \ref{subsec:ksdHawkes}, ksdHawkes formulation does not allow for a continuous imbalance covariate. As in \textcite{morariu2018state}, the covariate is thus discretized in five bins: $(-1, -0.6)$, $[-0.6, -0.2)$, $[-0.2, 0.2)$, $[0.2, 0.6)$ or $[0.6, 1)$. ksdHawkes models will also be tested with either $S1$ or $S2$ spread specifications. Standard Hawkes models benchmarks will be tested with various numbers of exponentials in the kernel, but as explained in Section \ref{subsec:ksdHawkes}, ksdHawkes models will only be tested with $d_n=1$. 

\subsection{Data}
\label{subsec:data}

We use tick-by-tick data provided by the Thomson-Reuters Tick History (TRTH) database for 36 stocks traded on the Paris stock exchange in 2015. The order flow is reconstructed with the method described in \textcite{munitoke2016reconstruction}. We only consider trading data between 10:00 and 14:00, as the volume of trading is not constant during the day, which makes the assumption of a constant baseline intensity unrealistic. This 4-hour interval length is a compromise between this assumption and the need for large samples. The timestamps are precise to the millisecond. On liquid stocks, one may often observed several orders submitted with less than a millisecond interval, which leads to events registered at the same timestamp. Such an event has probability 0 in a model with intensity-driven simple point processes. Such observations may lead to numerical instabilities and bias when estimating kernel parameters, as shown in a mini example in Appendix \ref{appendix:miniExample}. In the following estimations, if several events happen at the same timestamp, then only the last one of them is kept.

\subsection{Model calibration, model selection and goodness-of-fit}
\label{subsec:fit}

We test 'msd-$x$-$d_n$' models for $x\in\{\text{I,S1,S2,S3}\}$ and $d_n=1,\ldots,5$, 'msd-I-$y$-$d_n$' models for $y\in\{\text{S1,S2,S3}\}$ and $d_n=1,\ldots,5$, and benchmark them with 'ksd-$x$-$1$' models for $x\in\{\text{I,S1,S2}\}$' and 'std-$d_n$' models for $d_n=1,\ldots,5$, which amounts to 43 tested specifications.
Each specification is tested on each available trading day and each available stock in the sample, which amounts to more than 8000 stock-trading days.
Standard Hawkes models are fitted by standard maximum-likelihood estimation with explicit gradient computations. ksdHawkes models are fitted using the 'mpoints' library of \textcite{morariu2018state}. msdHawkes models are fitted using the maximum-likelihood procedure with explicit gradient computation described in Section \ref{sec:msdHawkes}. Expectation-Maximization algorithm is not used in these empirical tests because of its computational cost.

\begin{figure}
\includegraphics[width=\textwidth]{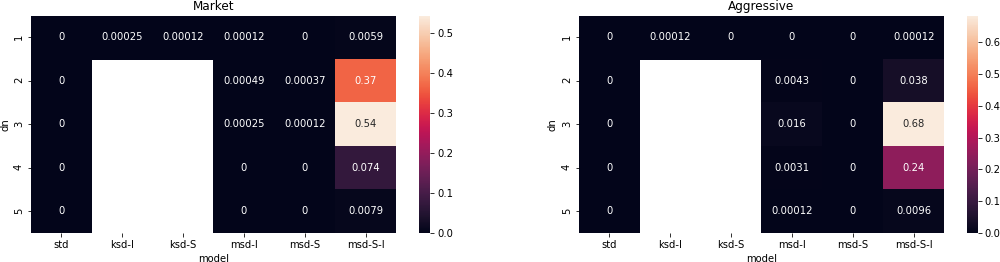}
\caption{Probability of selection of the tested models by AIC in the `Market' case (left) and the `Aggressive' case (right).}
\label{fig:msdHawkes-AIC}
\end{figure}

We use the Akaike information criterion (AIC) to select relevant models in these 43 specifications. Figure \ref{fig:msdHawkes-AIC} plots the probability that a given (group of) model is selected by AIC. In this context, 'msd-S' means a msdHawkes model with any of the three spread covariates.
Several observations can be made. Firstly, msdHawkes models are almost always selected by AIC. Benchmark models are not selected on the sample. Secondly, it appears clearly that msdHawkes models with a complete state-space representation with both spread and imbalance are largely favored by this criterion: 'msd-S-I' models are chosen more than 99\% of the stock-trading days in the `Market' case, and more than 97\% of the time in the 'Aggressive" case.
Thirdly, $d_n=3$ different timescales are favored in both cases. More precisely, AIC selects a 'msd-S-I' model with 2 or 3 exponentials in the Hawkes kernels 91\% of the of the stock-trading days in the `Market' case. In the 'Aggressive" case, AIC selects a 'msd-S-I' model with 3 or 4 exponential terms in the Hawkes kernels 92\% of the of the stock-trading days.

\begin{figure}
\centering
\includegraphics[width=0.8\textwidth]{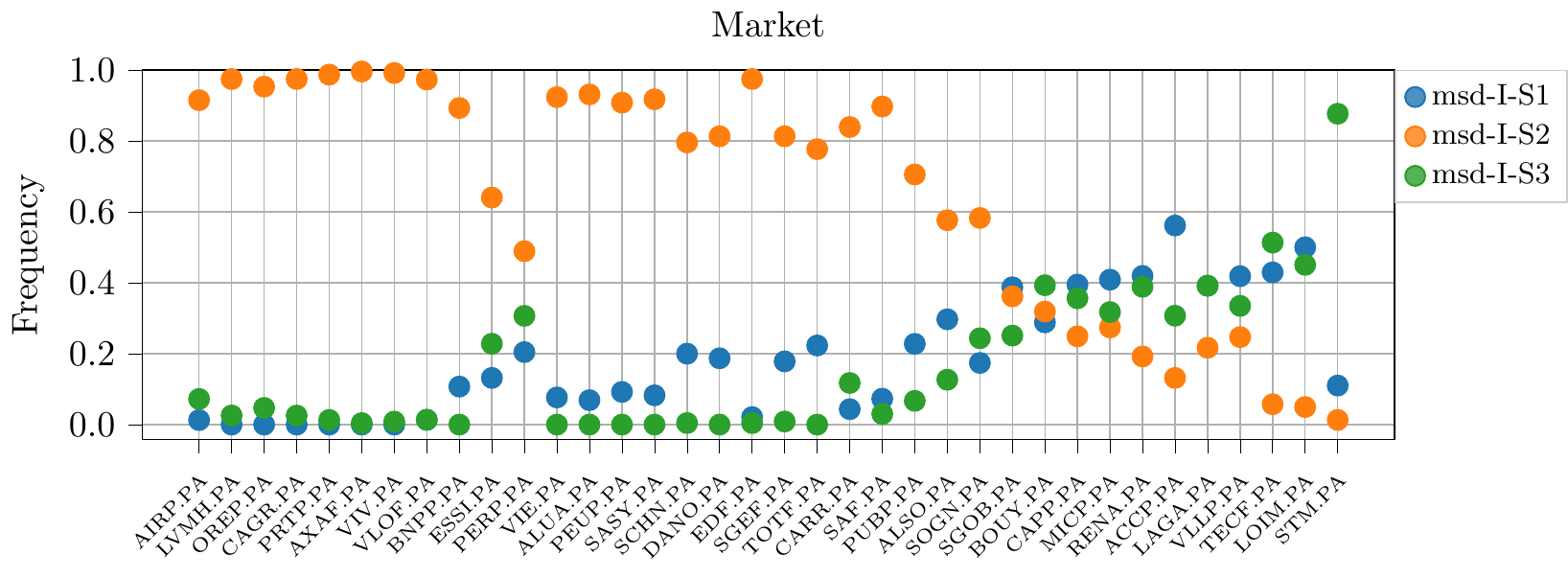}
\caption{Proportion of trading days where the AIC criterion chose a model with the S1, S2 and S3 specification in the state function (`Market' case).}
\label{fig:msdHawkes-AIC-SpreadMarket}
\end{figure}

We can refine the model selection by looking at the selected spread specification depending on the stock in the `Market' case. Figure \ref{fig:msdHawkes-AIC-SpreadMarket} plots the proportion of trading days where the AIC criterion selects a model with the S1, S2 and S3 specification in the state function (with any number of exponential terms in the kernel). Stocks are ordered by increasing average spread. 
It appears that for rather large-tick stocks (left part of the plot), models including spread in the S2 specification are largely selected, generally more than 80\% of the time. However when the average spread increases (small-tick stocks, right part of the graph), the probability of selecting the S2 specification regularly and significantly decreases. We do not have any similar observation in the `Aggressive' case, which might indicate that this specification impacts mostly market orders, and less limit orders and cancellations.

\begin{figure}
\includegraphics[width=\textwidth]{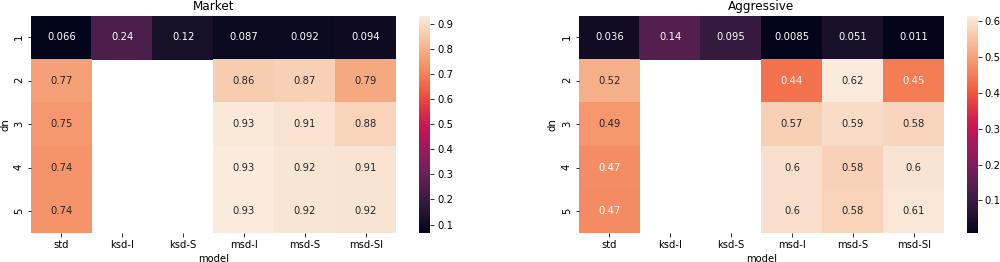}
\caption{Percentage of stock-trading days for which the residuals of a given model pass a Kolmogorov-Smirnov test at the 5\% level.}
\label{fig:msdHawkes-KSTests}
\end{figure}

We complement the AIC model selection by looking at the statistical significance of the tested model specifications. For a $d_e$-dimensional point process $(t_i^e)_{\substack{1 \leqslant e \leqslant d_e \\ i \in \mathbb{N}^*}}$ with intensity $\left(\lambda^e \right)_{1 \leqslant e \leqslant d_e}$, we introduce $d_e$ time series $r^1$, \dots, $r^{d_e}$, called residuals, and defined by
\begin{equation}
    \forall e \in \intervalleEntier{d_e}, \quad \forall i \in \mathbb{N}^*, \quad r^e_i = \int_{t_i^e}^{t_{i+1}^e} \lambda^e(t) \diff t.
\label{eq:residuals}
\end{equation}
Following \textcite{meyer1971}, each $(r^e_i)_{i \in \mathbb{N}^*}$ is i.i.d. and follows exponential law with parameter one. In the same way, we can compute the estimated residuals $\hat{r}^1$, \dots, $\hat{r}^{d_e}$ by replacing $\lambda^e$ by the estimated intensity $\hat{\lambda}^e$ in \eqref{eq:residuals}. All computational details for the msdHawkes residuals are given in Appendix \ref{appendix:Residuals}. To evaluate the quality of the fit, we check that the estimated residuals indeed follow an exponential law with parameter one. More precisely, we apply a Kolmogorov-Smirnov test at 95\% at each residuals sequence ($\hat{r}^e$ corresponds to the sequence of the $e$-th dimension for one trading day). A fit is considered validated for a trading day if the tests are passed on \emph{both} coordinates.
Figure \ref{fig:msdHawkes-KSTests} gives the percentage of days for which the fit is validated.
These results are in line with the AIC selection: $d_n=3$ exponential terms in the Hawkes kernels appear to be optimal for the statistical fit of both `Market' and `Aggressive' cases: a smaller number of exponential terms significantly degrades the performance, while adding more timescales in the model only marginally increases the number of stock-trading days for which a fit passes the KS tests. This stresses the importance of multiple timescales in Hawkes modeling \citep{lallouache2016limits}. It is also observed that state-dependency significantly increases the goodness-of-fit of the model compared to the standard (state-independent) Hawkes processes. Contrary to the AIC selection however, the different state functions (I, S, S-I) give similar goodness-of-fit performances.

\subsection{Estimated parameters}
\label{subsec:parameters}

\begin{figure}
\centering
\includegraphics[width=0.988\textwidth]{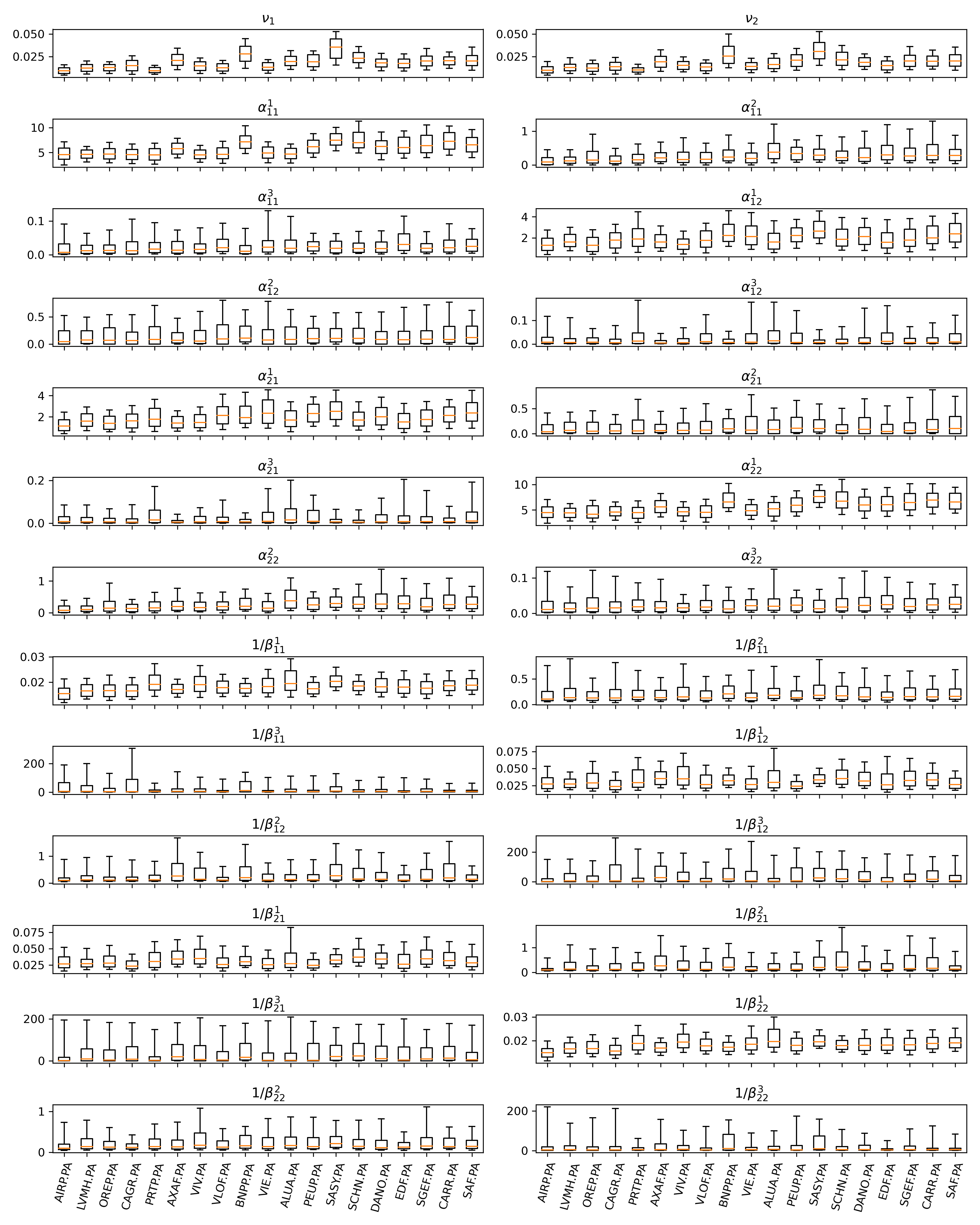}
\caption{Boxplots of estimated valued of the Hawkes parameters of the 'msd-I-S2-3' model for 19 large tick stocks traded in Paris in 2015 (`Market' case). The inverse of $\beta$'s are expressed in seconds.}
\label{fig:msdHawkes-EstimatedParams-hawkes}
\end{figure}

\begin{figure}
\centering
\includegraphics[width=\textwidth]{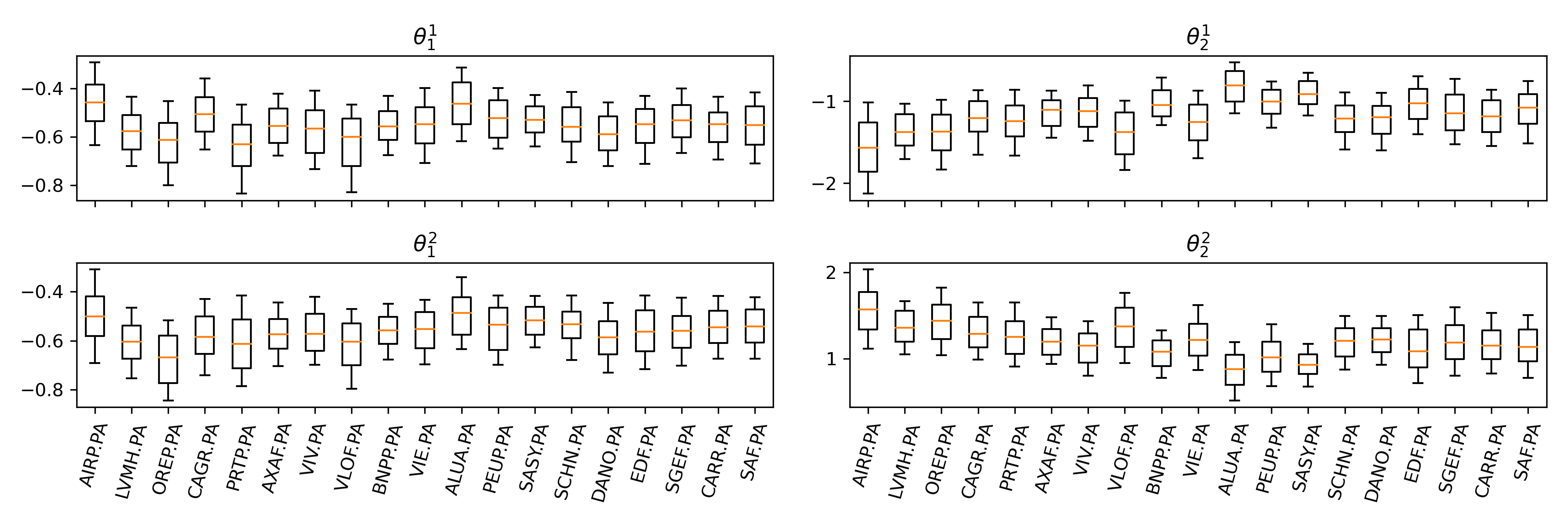}
\caption{Boxplots of estimated valued of the state parameters of the 'msd-I-S2-3' model for 19 large tick stocks traded in Paris in 2015 (`Market' case).}
\label{fig:msdHawkes-EstimatedParams-state}
\end{figure}

Although precise model selection might in the end be stock-dependent, results of the previous subsections show that general observations can nonetheless be made. Figures \ref{fig:msdHawkes-AIC} and \ref{fig:msdHawkes-KSTests} argue for models with both imbalance and spread, and $d_n=3$ exponential terms in Hawkes kernels, and figure \ref{fig:msdHawkes-AIC-SpreadMarket} identifies 19 stocks for which S2 is the best spread specification in terms of AIC.
We therefore study the fitted parameters of the 'msd-S2-I-3' model on these 19 stocks. Figure \ref{fig:msdHawkes-EstimatedParams-hawkes} provides the boxplots of the Hawkes parameters $\nu$, $\alpha$ and $\beta$, and Figure \ref{fig:msdHawkes-EstimatedParams-state} provides the boxplots of the state parameters $\theta$ for these 19 stocks.
Whiskers of the boxplots span the $[0.1,0.9]$ quantiles of the distributions, and outliers are not shown for easier readability. These figures show a very strong similarity of the estimated values of the parameters: for all parameter, the estimated values for all these stocks lie a similar range. We also observe a very strong symmetry in the cross- and self-excitation parameters of the Hawkes kernels: $\alpha_{12}^k$ and $\alpha_{21}^k$ share the same $y$-axis range for a given $k$, and so do $\alpha_{11}^k$ and $\alpha_{22}^k$, and so do the respective $\beta$'s. All stocks thus share the same three typical timescales in the Hawkes kernels. If we look at the characteristic time $\frac{1}{\beta_{ij}^k}$ of the $k$-th exponential term of the impact of an event of type $j$ on an event of type $i$, we observe that all stocks share typical timescales with the following order of magnitudes : a few seconds ($1/\beta_{ij}^3$), 100 milliseconds ($1/\beta_{ij}^2$) and 20 milliseconds ($1/\beta_{ij}^1$). More precisely, if we compute the mean of the characteristic times $1/\beta_{ij}^k$ across these 19 stocks, we obtain 18 ms, 142 ms, and 4.6 s for for the self-excitation kernels, and 30 ms, 136 ms and 5.2 s for the cross-excitation kernels (`Market' case). The exponential term $\beta_{ij}^3$ with the longest characteristic time exhibit however larger variations.

State parameters fitted by the msdHawkes model also exhibit stability with respect to time, stability with respect to stocks, and symmetry with respect to the bid/ask or up/down processes. Figure \ref{fig:msdHawkes-EstimatedParams-state} show for all stocks in the `Market' case that $\theta^1_1\approx\theta^2_1\approx -0.55$ for the spread coefficient, which means that when the spread is equal to one tick, the intensity of submission of bid and ask market orders is multiplied by approximately 3 in the model, all other things being equal. Similarly, comparing the top and bottom plots on the right column of Figure \ref{fig:msdHawkes-EstimatedParams-state} shows a striking symmetry in the imbalance parameters for all stocks, with $\theta^1_2\approx-\theta^2_2\approx -1.2$, which means for these stocks that the intensity of submission of ask (resp. bid) market orders is multiplied by approximately 2.5 when imbalance reaches 0.75 (resp. -0.75) compared to the case of equal bid and ask queues.

\subsection{Endogeneity}
\label{subsec:endogeneity}

\begin{figure}
\centering
\includegraphics[width=0.7\textwidth]{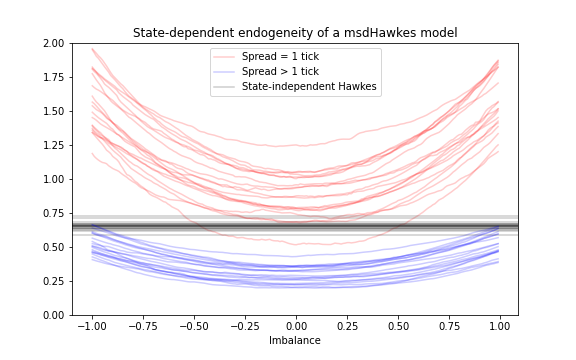}
\caption{State-dependent endogeneity in the 'msd-I-S2-3' model for 19 large tick stocks traded in Paris in 2015 (`Market' case).}
\label{fig:msdHawkes-Endogeneity}
\end{figure}

In a standard Hawkes model, the endogeneity coefficient is defined as the spectral radius of the matrix of the L1-norm of the excitation kernels. In the case of a $d_e$-dimensional process with kernels with $d_n$ exponential terms, this is the spectral radius of the matrix 
$\left(\sum_{n=1}^{d_n}\frac{\alpha_{ij}^n}{\beta_{ij}^n}\right)_{i,j=1,\ldots,d_e}
$.
It is called endogeneity because in a branching structure representation of the Hawkes process, it represents the fraction of events triggered by previous events (as opposed to immigrants, i.e. exogenous events triggered by the baseline intensities). 
In the msdHawkes process, if we fix the state variables $(x_j)_{j=1,\ldots,d_x}$ and set $m_i=\exp\left(\sum_{j=1}^{d_x}\theta^i_j x_j\right)$, then we have a standard Hawkes model with baseline intensity $m_i\nu_i$ and Hawkes coefficients $(m_i\alpha_{ij}^n,\beta_{ij}^n)$, $i,j=1,\ldots,d_e$, $n=1,\ldots,d_n$. We can therefore compute a state-dependent endogeneity coefficient. 
Figure \ref{fig:msdHawkes-Endogeneity} plots the median state-dependent endogeneity coefficient in the 'msd-I-S2-3' model for the 19 stocks selected in Section \ref{subsec:parameters}.
For each stock, three endogeneity curves are plotted as a function of the imbalance: one when the spread is equal to 1 tick (in red), one when the spread is greater than 1 tick (in blue), and one for the standard (non-state-dependent) reference (in black). It is remarkable to observe the agreement of the 3 groups of 19 curves in this group of stocks. In the 'msd-I-S2-3' model, we clearly identify several regimes. When the spread is greater than one tick (blue lines), the endogeneity coefficient is below the endogeneity of a standard Hawkes model 'std-3' (black lines). It increases when the imbalance gets away from 0, but slightly. When the spread is equal to one tick however, the coefficient rises above the standard Hawkes reference, and has more steep branches for the dependence on the imbalance. The coefficient may even rise above the critical value 1, indicating that the msdHakwes models temporarily explores regimes that would not be stable if they were coefficients of standard Hawkes processes. It is interesting to observe that these results are very much in line with the results of \textcite{morariu2018state} in which similar endogeneity proxies are computed for the ksdHawkes models, for either the imbalance or the spread dependency. The msdHawkes framework provides here the full two-dimensional dependency pictures in agreement with the previous one-dimensional dependence models.

\subsection{Out-of-sample prediction}
\label{sec:prediction}

In this section we test the msdHawkes models on a prediction exercise used in \textcite{munitoke2020analyzing}. The (theoretical) exercise consists in predicting at any time $t$ the type of the incoming event, if one were to occur right now. In an intensity-based model, the predicted type is simply the coordinate of the point process with the highest intensity at $t-$.
In other words, assume that an order of a certain type $e$ (bid or ask in the `Market' case, up or down in the `Aggressive' case) arrives on the market at time $t$. Right before, at $t-$, an intensity-based point process model would have predicted that the type of the upcoming order would be $\argmax_{e' \in \intervalleEntier{d_e}} \lambda^{e'}(t-)$ (the one with the biggest intensity). If this prediction is equal to $e$, it is correct. For every trading day, we make our prediction based on the parameters estimated on \emph{the previous trading day} (and therefore out-of-sample).
Observe that this is not an exact exercise with perfect benchmark: if we consider a two-dimensional homogeneous Poisson process with independent coordinates with respective intensities $\lambda_1=1$ and $\lambda_2=2$, then the full knowledge of the model leads to always predict $2$, with a percentage of success equal to $\frac{2}{3}$.

\begin{figure}
\centering
\includegraphics[width=0.6\textwidth]{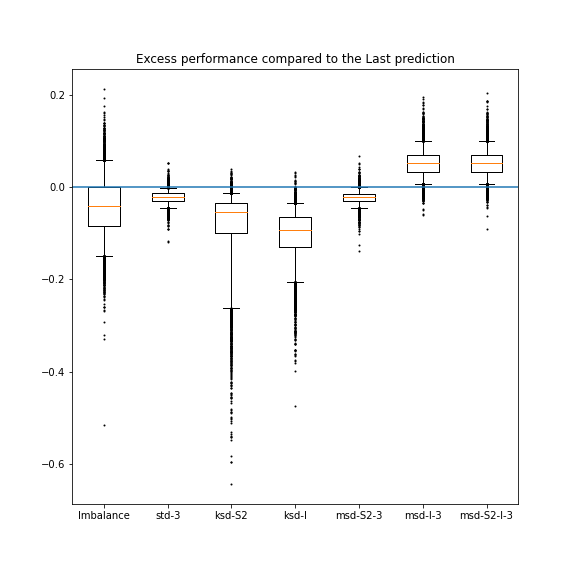}
\caption{Excess performances for several model-based predictions of the next trade sign compared to the 'Last' prediction, for 36 stocks traded in Paris in 2015 (`Market' case).}
\label{fig:msdHawkes-Prediction}
\end{figure}

The msdHawkes prediction is compared to several benchmarks. The 'Last' method always predicts that the incoming order is of the same type as the previous order. The 'Imbalance' method predicts a bid order (resp. downwards movement) in the `Market' (resp. `Aggressive') case if the imbalance is negative, and symmetric predictions if it is positive. For completeness, we also use as comparison the 'ksd-I' and 'ksd-S2' models, but bear in mind that these models are not designed for such an exercise (see discussion in Section \ref{subsec:ksdHawkes}).
We run this prediction exercise on the 36 stocks available in the sample, which amounts to more the 8000 stock-trading days. Since it has been suggested in \textcite{munitoke2020analyzing} that parsimonious Hawkes models without cross-excitation perform better in this prediction exercise, the stdHawkes and msdHawkes models used in this section have no cross-excitation term in the kernel (and $d_n=3$ as in the previous sections).
Figure \ref{fig:msdHawkes-Prediction} plots for the `Market' case the \emph{excess} accuracy of several models compared to the 'Last' benchmark (horizontal blue line at $0$). 
In this case, the 'Last' benchmark correctly signs in average $81\%$ of the trades (in this exercise, we keep the events occurring at the same millisecond: this has an effect on the absolute accuracy of the methods, but this does not change the relative performances of the methods, hence the analysis). This is the best of the benchmarks: 'Imbalance' is outperformed on 75\% of the stock-trading days, the standard Hawkes model and the ksdHawkes models are outperformed on more than 90\% of the sample. The msdHawkes model with only a spread dependency has a performance very similar to the standard Hawkes model: indeed, the spread being a symmetric variable for the bid and the ask (see Section \ref{subsec:parameters}), the bid and ask $\theta$'s associated to the spread are very close, and thus do no provide any information on the signature. The msdHawkes model with the imbalance however significantly improves the signature. Excess performance is positive 97.5\% of the time, and increases the signature accuracy by more than 5\% in average on the sample. Actually, the msd-I models act like a combination of the 'Last' and 'Imbalance' predictors in this exercise: the information carried by the first one is caught by the Hawkes factor of the intensity, while the information given by the second one is included in the state factor. As discussed, adding a spread covariate neither improves nor degrades the performance. 

\begin{rmk}
We can run a similar exercise in the `Aggressive' case. Imbalance is in this case the best predictor here, with an average accuracy close to 79 \%. Ranking the models by accuracy yields results similar to the `Market' case: msdHawkes models with an imbalance state covariate performs much better (roughly +20\%) than standard Hawkes, ksdHawkes models with spread or imbalance, or msdHawkes with spread. However, they do not improve the basic imbalance benchmark (-2\% in accuracy in average).
\end{rmk}

\section{Conclusion}

We have proposed a Hawkes-based state-dependent point process to model order flows in limit order books. Compared to other state-dependent Hawkes-based models, this formulation provides interesting modeling features in terms of parsimony, as well as improved performances in terms of goodness-of-fit to empirical data for example. Empirical results provide insights on microstructure mechanisms in the context a general state process including imbalance and spread covariates (state-dependent endogeneity). This contribution calls for further improvements in extending the standard Hawkes process to state-dependent microstructure modeling. Future works could take several directions. One of them is to statistically determine which covariates should be included in a LOB state representation, besides the spread and imbalance used here. Results might be stock-dependent, or characterize the microstructure of different stock groups, as suggested in the results presented here. In another direction, these works in empirical finance could trigger several mathematical works to better understand the probabilistic and statistic properties of such processes, designed at the moment for empirical financial modeling.

\section*{Acknowledgements}
The authors thank the reviewers for their useful comments. During the accomplishment of this work Ioane Muni Toke has been partially supported by the Japan Science and Technology Agency (Grant number CREST JPMJCR2115).

\printbibliography

\appendix
\numberwithin{equation}{section}

\section{Maximum-likelihood estimation of msdHawkes processes}
\label{appendix:MLE}

\subsection{Log-likelihood computation}
\label{appendix:LogLik}
This section provides efficient (recursive, linear) computations for the log-likelihood of msdHawkes processes and its gradients. These computations allow for a fast (linear in the length of the sample) maximum likelihood estimation of these processes. We use the notations introduced in Section \ref{sec:msdHawkes}.
Let $e \in \intervalleEntier{d_e}$. For a given sample, let let $n_e$ be the number of events of type $e$ and $\mathcal T^e=\{t^e_i:i=1,\ldots,n_e\}$ be the set of events of type $e$. Maximum-likelihood estimation requires the maximization of the function
\begin{equation}
    \begin{split}
        L^e_T(\nu_e, \alpha_{e\cdot}, \beta_{e\cdot}, \theta^e) = &- \int_0^T \lambda^{H,e} (s) e^{\left< X_s, \theta^e \right>}\diff s + \int_0^T \log(\lambda^{H,e} (s))\diff N^e_s\\ 
        &+ \int_0^T \left< X_{s-}, \theta^e \right> \diff N^e_s.
    \end{split}
\label{eq:loglikapp}
\end{equation}
The third term of Equation \eqref{eq:loglikapp} is straightforward to compute. The second term of Equation \eqref{eq:loglikapp} can be written:
\begin{equation}
    \sum_{i=1}^{n_e} \log\left( \nu_e + \sum_{e'=1}^{d_e}\sum_{n=1}^{d_n}\alpha_{ee'}^n R^{ee'n}(i)\right),
\end{equation}
where the quantity $R^{ee'n}(i)$, $e' \in \intervalleEntier{d_e}$, $n \in \intervalleEntier{d_n}$, $i \in \intervalleEntier{n_e}$, is by definition: 
\begin{equation}
    R^{ee'n}(i) = \int_{]0,t_i^e[} e^{-\beta_{ee'}^n(t_i^e-u)}\diff N_u^{e'} =
\sum_{\substack{t_j^{e'} \in \mathcal{T}^{e'} \\ t_j^{e'} < t_i^{e}}} e^{-\beta_{ee'}^n(t_i^e - t_j^{e'})},
\end{equation}
and can be computed recursively in the following way:
\begin{equation}
\left\{\begin{array}{rl}
R^{ee'n}(1) & = \sum_{\substack{t_j^{e'} \in \mathcal{T}^{e'} \\ t_j^{e'} < t_0^{e}}} e^{-\beta_{ee'}^n(t_0^e - t_j^{e'})}  \quad \quad \quad\left(=0 \text{ if }e'=e \right), 
\\
R^{ee'n}(i) & = e^{-\beta_{ee'}^n(t_i^e-t_{i-1}^e)}R^{ee'n}(i-1) + \sum_{\substack{t_j^{e'} \in \mathcal{T}^{e'} \\ t_{i-1}^{e} \leqslant t_j^{e'} < t_i^{e}}} e^{-\beta_{ee'}^n(t_i^e - t_j^{e'})},
\qquad\qquad (i \in \intervalleEntier[2]{n_e}).
\end{array}\right.
\end{equation}

We can develop as well recursive computations for the first term of Equation \eqref{eq:loglikapp}. Since we assume that $X$ is piecewise continuous, we have:
\begin{equation}
    X = \sum_{i = 0}^{N-1} x_i \mathbf{1}_{[\tau_i, \tau_{i+1}[},
\end{equation}
where $\tau_0 = 0$ and $\tau_N = T$. Without any loss of generality, we now add the jump times of $N$ to the vector $(\tau_0, ..., \tau_N)$, and renumber it. The first term of Equation \eqref{eq:loglikapp} is then written:
\begin{equation}
\begin{split}
\int_0^T \lambda^H(s)e^{\left< X_s, \theta \right>}\diff s = &\nu_e \sum_{j = 0}^{N-1}  e^{\left<x_j,\theta^e \right>}\left(\tau_{j+1} - \tau_{j}\right) \\
&+ \sum_{e' = 1}^{d_e}\sum_{n = 1}^{d_n}\sum_{j=1}^{N-1} \alpha_{ee'}^ne^{\left<x_j,\theta^e \right>} \int_{\tau_{j}}^{\tau_{j+1}}\left(\sum_{\substack{t_i^{e'} \in \mathcal{T}^{e'} \\ t_i^{e'} < s}} e^{-\beta_{ee'}^n (s-t_i^{e'})}\right)\diff s .
\end{split}
\label{eq:loglikapp-term1}
\end{equation}
The first term of the above equation \eqref{eq:loglikapp-term1} is straightforward to implement numerically. After some further computations, we show that the second term of the above equation \eqref{eq:loglikapp-term1} satisfies:
\begin{equation}
I_e = \sum_{n = 1}^{d_n}\sum_{e' = 1}^{d_e}\sum_{j=1}^{N-1} \alpha_{ee'}^ne^{\left<x_j,\theta^e \right>} \int_{\tau_{j}}^{\tau_{j+1}}\left(\sum_{\substack{t_j^{e'} \in \mathcal{T}^{e'} \\ t_j^{e'} < s}} e^{-\beta^n_{ee'} (s-t_i^{e'})}\right)\diff s
= \sum_{n=1}^{d_n}\sum_{e' = 1}^{d_e}\frac{\alpha_{ee'}^n}{\beta_{ee'}^n}\sum_{i=1}^{n_{e'}} S^{ee'n}(i),
\end{equation}
where $t_{n_{e'}+1}^{e'} = T$ and the $S^{ee'n}(i)$'s satisfy the backwards recursive formulation: $S^{ee'n}(n_{e'} + 1) = 0$ and for $i\in \intervalleEntier{n_{e'}}$:
\begin{equation}
     S^{ee'n}(i) = e^{-\beta_{ee'}^n (t_{i+1}^{e'}-t_i^e)}S^{ee'}(i + 1) + \sum_{\substack{j \in \lbrace 1,...,N-1 \rbrace \\ t_i^{e'} \leqslant \tau_j < t_{i+1}^{e'}}}e^{\left<x_j,\theta^e \right>}e^{-\beta_{ee'}^n\left( \tau_j - t_i^{e'} \right)}\left(1 - e^{-\beta_{ee'}^n\left(\tau_{j+1} - \tau_j\right)}\right).
\end{equation}

\subsection{Computationally efficient gradient formulations}
\label{appendix:gradLogLik}

In this section we provide computational forms of the partial derivatives of the log-likelihood given by equation \eqref{eq:loglikapp}, to be used in maximum likelihood estimation. In order to avoid unnecessarily cumbersome notations, we do not write the dependencies of the functions to the parameters. We keep the notations of Section \ref{sec:msdHawkes} and Appendix \ref{appendix:LogLik}.

Gradients w.r.t. the baseline intensities and the $\alpha$'s are straightforwardly computed. Since the coefficients $R^{ee'}(i)$ do not depend on $\nu_e$, a straightforward calculation gives the partial derivative w.r.t. the baseline intensities:
\begin{equation}
    \dpart{L^e_T}[\nu_e] = - \sum_{j = 0}^{N-1}  e^{\left<x_j,\theta^e \right>}\left(\tau_{j+1} - \tau_{j}\right) + 
    \sum_{i=1}^{n_e} \frac{1}{ \nu_e + \sum_{e'=1}^{d_e}\sum_{n=1}^{d_n}\alpha_{ee'}^n R^{ee'n}(i)}.
\end{equation}
As the coefficients $R^{ee'n}(i)$ and $S^{ee'n}(i)$ do not depend on $\alpha_{ee'}$, we have:
\begin{equation}
    \dpart{L^e_T}[\alpha_{ee'}^n] = -\frac{1}{\beta_{ee'}^n}\sum_{i=1}^{n_{e'}} S^{ee'n}(i) + 
    \sum_{i=1}^{n_e} \frac{R^{ee'n}(i)}{ \nu_e + \sum_{e''=1}^{d_e}\sum_{n'=1}^{d_n}\alpha_{ee''}^n R^{ee''n'}(i)}.
\end{equation}

Gradients for the state coefficients are a bit more involved. We have:
\begin{equation}
    \nabla_{\theta^e}L^e_T = - \nu_e  \sum_{j = 0}^{N-1}  x_j e^{\left<x_j,\theta^e \right>}\left(\tau_{j+1} - \tau_{j}\right) - \nabla_{\theta^e} I^e + \sum_{i=1}^{n_e} X(t_i^e),
\end{equation}
which can be recursively computed. Indeed, defining the notation $S^{ee'n}_{\theta}(i) := \nabla_{\theta^e}S^{ee'n}(i)$, the middle term is written:
\begin{equation}
    \nabla_{\theta^e}I^e = \sum_{e' = 1}^{d_e}\sum_{n=1}^{d_n}\frac{\alpha_{ee'}^n}{\beta_{ee'}^n}\sum_{i=1}^{n_{e'}} S^{ee'n}_{\theta}(i),
\end{equation}
where the coefficients $S^{ee'n}_{\theta}(i)$ can be computed recursively $S^{ee'n}_{\theta}(n_{e'} + 1) = 0$, $t_{n_{e'}+1}^{e'} = T$, and for all $i \in \intervalleEntier{n_{e'}}$: 
\begin{equation}
\begin{split}
    S^{ee'n}_{\theta}(i) = &e^{-\beta_{ee'}^n (t_{i+1}^{e'}-t_i^e)}S^{ee'}_{\theta}(i + 1)\\
    &+\sum_{\substack{j \in \lbrace 1,...,N-1 \rbrace \\ t_i^{e'} \leqslant \tau_j < t_{i+1}^{e'}}}
    x_j e^{\left<x_j,\theta^e \right>}e^{-\beta_{ee'}^n\left( \tau_j - t_i^{e'} \right)}\left(1 - e^{-\beta_{ee'}^n\left(\tau_{j+1} - \tau_j\right)}\right).
\end{split}
\end{equation}

We finally turn to the partial derivatives w.r.t. the $\beta$'s. Again, let $e' \in \intervalleEntier{d_e}$ and $n \in \intervalleEntier{d_n}$.
We define $S^{ee'n}_{\beta}(i) := \dpart{S^{ee'n}}[\beta_{ee'}^n](i)$ and $R^{ee'n}_{\beta}(i) :=
\dpart{R^{ee'n}}[\beta_{ee'}^n](i)$. Using these notations, we have:
\begin{equation}
\begin{split}
    \dpart{L^e_T}[\beta_{ee'}^n] = 
    - &\frac{\alpha_{ee'}^n}{\beta_{ee'}}\sum_{i=1}^{n_{e'}} \left( S^{ee'n}_{\beta}(i) - \frac{1}{\beta_{ee'}^n}S^{ee'n}(i)\right)\\
    &+ \sum_{i=1}^{n_e} \frac{\alpha_{ee'}^n R^{ee'n}_{\beta}(i)}{ \nu_e + \sum_{e''=1}^{d_e}\sum_{n'=1}^{d_n}\alpha_{ee''}^n R^{ee''n'}(i)}.
\end{split}
\end{equation}
After further computations, we show that the coefficients $R^{ee'n}_{\beta}(i)$ satisfy the recursive formulation:
\begin{equation}
    \left\{\begin{array}{rl}
R^{ee'n}_{\beta}(1) & = \sum_{\substack{t_j^{e'} \in \mathcal{T}^{e'} \\ t_j^{e'} < t_0^{e}}} 
 -(t_0^e - t_j^{e'})e^{-\beta_{ee'}^n(t_0^e - t_j^{e'})},
\\
   R^{ee'}_{\beta}(i) & = e^{-\beta_{ee'}^n(t_i^e-t_{i-1}^e)}
\left(R^{ee'n}_{\beta}(i-1) - (t_i^{e}-t_{i-1}^e)R^{ee'n}(i-1) \right)
\\
& \qquad\qquad - \sum\limits_{\substack{t_j^{e'} \in \mathcal{T}^{e'} \\ t_{i-1}^{e} \leqslant t_j^{e'} < t_i^{e}}} 
(t_i^e - t_j^{e'})e^{-\beta_{ee'}^n(t_i^e - t_j^{e'})}, \qquad\qquad (i \in \intervalleEntier[2]{n_{e}}) .
\end{array}\right.
\end{equation}
Then by differentiating, we get:
\begin{equation}
    S^{ee'n}_{\beta}(i) =
    \sum_{\substack{j \in \lbrace 1,...,N-1 \rbrace \\ t_i^{e'} \leqslant \tau_j}}
    \left(-\left(\tau_{j} - t_i^{e'}\right)e^{-\beta_{ee'}^n\left(\tau_{j} - t_i^{e'}\right)} 
    + \left(\tau_{j+1} - t_i^{e'}\right)e^{-\beta_{ee'}^n\left(\tau_{j+1} - t_i^{e'}\right)}\right).
\end{equation}
Setting $S^{ee'n}_{\beta}(n_{e'} + 1) = 0$ and $t_{n_{e'}+1}^{e'} = T$, we obtain the following recursion formula:
\begin{equation}
\begin{split}
    S^{ee'n}_{\beta}(i) = & e^{-\beta_{ee'}^n(t_{i+1}^{e'}-t_i^{e'})}
    \left( S^{ee'n}_{\beta}(i+1)  - (t_{i+1} - t_i)S^{ee'n}(i+1)\right)\\
     & + \sum_{\substack{j \in \lbrace 1,...,N-1 \rbrace \\ t_i^{e'} \leqslant \tau_j < t_{i+1}^{e'}}}
    x_j e^{\left<x_j,\theta^e \right>}e^{-\beta_{ee'}^n\left( \tau_j - t_i^{e'} \right)}\left(- \left( \tau_j - t_i^{e'} \right) + \left(\tau_{j+1} - \tau_j\right)
    e^{-\beta_{ee'}^n\left(\tau_{j+1} - \tau_j\right)}\right).
\end{split}
\end{equation}
All these results allow for a computationally efficient implementation of the maximum likelihood estimation of msdHawkes processes.

\section{EM algorithm for msdHawkes processes}
\label{appendix:EM_algo}

In this section we provide the full computations and EM algorithm for the estimation of msdHawkes processes.
The expected complete log-likelihood for the coordinate $e=1,\ldots,d_e$ is:
\begin{align}
\mathbf E\left[L^{c,e}_T(\nu,\alpha,\beta,\theta)\right]
= & 
-\int_0^T \nu_e e^{\langle \theta^e,X_{t} \rangle}\,\diff t
+ \sum_{t^{e}_i} \langle \theta^e,X_{t^e_i-}\rangle
+ \sum_{t^{e}_i} \mathbf P(u^{e}_i=(e,i)) \log \nu_e
\nonumber \\ & 
+ \sum_{e'=1}^{d_e} \sum_{t^{e'}_j} \sum_{n'=1}^{d_n} \Bigg[
-\int_{t^{e'}_j}^{T} \alpha_{ee'}^{n'} e^{-\beta_{ee'}^{n'}(t-t^{e'}_j)} e^{\langle \theta^e,X_{t} \rangle}\,\diff t
\nonumber \\ & 
\quad + \sum_{t^{e}_i : t^{e'}_j<t^{e}_i} \mathbf P(u^{e}_i=(e',n',j)) \log\left(\alpha_{ee'}^{n'} e^{-\beta_{ee'}^{n'} (t^e_i-t^{e'}_j)}  \right)
\Bigg] .
\label{eq:ethECLL-full-dnsplit}
\end{align}
With some computations, we show that:
\begin{equation}
 - \sum_{e'=1}^{d_e} \sum_{t^{e'}_j} \int_{t^{e'}_j}^{T} \sum_{n'=1}^{d_n} \alpha_{ee'}^{n'} e^{-\beta_{ee'}^{n'}(t-t^{e'}_j)} e^{\langle \theta^e,X_{t} \rangle}\,dt
=
- \sum_{e'=1}^{d_e} \sum_{n'=1}^{d_n} \frac{\alpha_{ee'}^{n'}}{\beta_{ee'}^{n'}}  \sum_{t^{e'}_j} A^{n'}_{ee'}(j) 
\end{equation}
where
$A^{n'}_{ee'}(j) = \sum_{\tau_k: t^{e'}_j \leq \tau_k} e^{\langle \theta^e,X_{\tau_k+} \rangle} \left(e^{-\beta_{ee'}^{n'}(\tau_{k}-t^{e'}_j)}-e^{-\beta_{ee'}^{n'}(\tau_{k+1}-t^{e'}_j)}\right)$, and the $\tau_k$'s have been defined in Appendix \ref{appendix:LogLik}.
We can compute these coefficients $A^{n'}_{ee'}(j)$ with the backwards recursion:
\begin{equation}
A^{n'}_{ee'}(n^{e'}_j) = \sum_{\tau_k: t^{e'}_{n^{e'}_j} \leq \tau_k} e^{\langle \theta^e,X_{\tau_k+} \rangle} \left(e^{-\beta_{ee'}^{n'}(\tau_{k}-t^{e'}_{n^{e'}_j})}-e^{-\beta_{ee'}^{n'}(\tau_{k+1}-t^{e'}_{n^{e'}_j})}\right)
\end{equation}
and
\begin{equation}
A^{n'}_{ee'}(j-1)
= \sum_{\tau_k: t^{e'}_{j-1} \leq \tau_k < t^{e'}_{j}} e^{\langle \theta^e,X_{\tau_k+} \rangle} \left(e^{-\beta_{ee'}^{n'}(\tau_{k}-t^{e'}_{j-1})}-e^{-\beta_{ee'}^{n'}(\tau_{k+1}-t^{e'}_{j-1})}\right)
+ e^{-\beta_{ee'}^{n'}(t^{e'}_{j}-t^{e'}_{j-1})}
A^{n'}_{ee'}(j).
\end{equation}

The partial derivative with respect to $\nu_e$ is written:
\begin{align}
\frac{\partial \mathbf E\left[L^e_T(\nu,\alpha,\beta,\theta)\right]}{\partial \nu_e} = -\int_0^T e^{\langle \theta^e,X_{t} \rangle}\,\diff t + \sum_{t^{e}_i} \mathbf P(u^{e}_i=(e,i)) \frac{1}{\nu_e},
\end{align}
and this derivative is zero if the following holds:
\begin{equation}
	\nu_e = \frac{ \sum_{t^{e}_i} \mathbf P(u^{e}_i=(e,i))}{\int_0^T e^{\langle \theta^e,X_{t} \rangle}\,\diff t}.
\end{equation}
The partial derivative with respect to $\alpha^{n'}_{ee'}$ is written:
\begin{align}
\frac{\partial \mathbf E\left[L^e_T(\nu,\alpha,\beta,\theta)\right]}{\partial \alpha^{n'}_{ee'}} = &
- \frac{1}{\beta^{n'}_{ee'}} \sum_{t^{e'}_j} A^{n'}_{ee'}(j)
+ \sum_{t^{e'}_j} \sum_{t^{e}_i : t^{e'}_j<t^{e}_i} \mathbf P(u^{e}_i=(e',n',j)) \frac{1}{\alpha^{n'}_{ee'}} ,
\end{align}
and this derivative is zero if the following holds:
\begin{equation}
\frac{\alpha^{n'}_{ee'}}{\beta^{n'}_{ee'}} = \frac{\sum_{t^{e'}_j} \sum_{t^{e}_i : t^{e'}_j<t^{e}_i} \mathbf P(u^{e}_i=(e',n',j))}{ \sum_{t^{e'}_j} A^{n'}_{ee'}(j)} .
\end{equation}
The partial derivative with respect to $\beta^{n'}_{ee'}$ is written:
\begin{align}
\frac{\partial \mathbf E\left[L^e_T(\nu,\alpha,\beta,\theta)\right]}{\partial \beta^{n'}_{ee'}} = &
\frac{\partial }{\partial \beta^{n'}_{ee'}} \sum_{t^{e'}_j} \Bigg[
-\int_{t^{e'}_j}^{T} \alpha_{ee'}^{n'} e^{-\beta_{ee'}^{n'}(t-t^{e'}_j)} e^{\langle \theta^e,X_{t} \rangle}\,\diff t
\nonumber \\ & 
\quad + \sum_{t^{e}_i : t^{e'}_j<t^{e}_i} \mathbf P(u^{e}_i=(e',n',j)) \log\left(\alpha_{ee'}^{n'} e^{-\beta_{ee'}^{n'} (t^e_i-t^{e'}_j)}  \right)
\Bigg].
\end{align}
Observe that:
\begin{align}
& \frac{\partial }{\partial \beta^{n'}_{ee'}} \sum_{t^{e}_i : t^{e'}_j<t^{e}_i} \mathbf P(u^{e}_i=(e',n',j)) \log\left(\alpha_{ee'}^{n'} e^{-\beta_{ee'}^{n'} (t^e_i-t^{e'}_j)}  \right)
\nonumber \\ = &
- \sum_{t^{e}_i : t^{e'}_j<t^{e}_i} \mathbf P(u^{e}_i=(e',n',j)) \, (t^{e}_i-t^{e'}_j),
\end{align}
and that:
\begin{align}
& - \frac{\partial }{\partial \beta^{n'}_{ee'}} \sum_{t^{e'}_j} \int_{t^{e'}_j}^{T} \alpha_{ee'}^{n'} e^{-\beta_{ee'}^{n'}(t-t^{e'}_j)} e^{\langle \theta^e,X_{t} \rangle}\,\diff t
\nonumber \\ = &
-\frac{\partial }{\partial \beta^{n'}_{ee'}} \sum_{t^{e'}_j} \sum_{\tau_k: t^{e'}_j \leq \tau_k}  \int_{\tau_k}^{\tau_{k+1}} \alpha^{n'}_{ee'} e^{-\beta^{n'}_{ee'}(t-t^{e'}_j)} e^{\langle \theta^e,X_{t} \rangle}\,\diff t
\nonumber \\ = &
\frac{\alpha^{n'}_{ee'}}{(\beta^{n'}_{ee'})^2} \sum_{t^{e'}_j} A^{n'}_{ee'}(j)
\nonumber \\ & \quad
+ \frac{\alpha^{n'}_{ee'}}{\beta^{n'}_{ee'}} \sum_{t^{e'}_j} \sum_{\tau_k: t^{e'}_j \leq \tau_k} e^{\langle \theta^e,X_{\tau_k+} \rangle} \left((\tau_{k}-t^{e'}_{j})e^{-\beta^{n'}_{ee'}(\tau_{k}-t^{e'}_{j})}-(\tau_{k+1}-t^{e'}_{j})e^{-\beta^{n'}_{ee'}(\tau_{k+1}-t^{e'}_{j})}\right)
,
\end{align}
so the derivative is zero if the following holds:
\begin{align}
0 = & \frac{1}{\beta^{n'}_{ee'}} \sum_{t^{e'}_j} A^{n'}_{ee'}(j)
\nonumber \\ & 
+ \sum_{t^{e'}_j} \sum_{\tau_k: t^{e'}_j \leq \tau_k} e^{\langle \theta^e,X_{\tau_k+} \rangle} \left((\tau_{k}-t^{e'}_{j})e^{-\beta^{n'}_{ee'}(\tau_{k}-t^{e'}_{j})}-(\tau_{k+1}-t^{e'}_{j})e^{-\beta^{n'}_{ee'}(\tau_{k+1}-t^{e'}_{j})}\right)
\nonumber \\ & 
- \left(\frac{\alpha^{n'}_{ee'}}{\beta^{n'}_{ee'}}\right)^{-1} \sum_{t^{e'}_j} \sum_{t^{e}_i : t^{e'}_j<t^{e}_i} \mathbf P(u^{e}_i=(e',n',j)) \, (t^{e}_i-t^{e',}_j).
\label{eq:beta_update_equation}
\end{align}

An EM-type algorithm can thus be written. Let $N=\sum_{e=1}^{d_e} n_e$ the total number of events. For $e=1,\ldots,d_e$:
\begin{enumerate}
	\item Initialize $\hat\nu_e^{(0)}$,$(\hat\alpha_{e,e'}^{n,(0)})_{e=1,\ldots,d_e, n=1,\ldots,d_n}$, $(\hat\beta_{e,e'}^{n,(0)})_{e=1,\ldots,d_e, n=1,\ldots,d_n}$, $(\hat\theta^{e,(0)}_j)_{j=1,\ldots,dx}$ and set $k=0$.
	\item Compute branching probabilities with estimated parameters of rank $(k)$ : for $i=1,\ldots,N$, $j=1,\ldots,i-1$,
\begin{equation}
\begin{cases}
	p^{(k)}_{i,i} = \frac{\nu_e}{\lambda^{H,e}(t^e_i-)},
	\\
	p^{(k)}_{i,j,n'} = \frac{\hat\alpha_{ee'}^{n'} e^{-\hat\beta_{ee'}^{n'}(t^e_i-t^{e'}_j)}}{\lambda^{H,e}(t^e_i-)}.
\end{cases}
\end{equation}
	\item Update state parameters $\theta^{e,(k+1)}$ by maximizing $\mathbf E\left[L^e_T(\nu,\alpha,\beta,\theta)\right]$ of Equation \eqref{eq:ethECLL-full-dnsplit} given the other parameters of rank ($k$) and the branching probabilities.
	\item Update Hawkes parameters with
	\begin{equation}
	\hat\nu_e^{(k+1)} = \frac{\sum_{t^e_i} p^{(k)}_{i,i}}{\int_0^T e^{\langle \theta^{e,(k)},X_{t} \rangle}\,\diff t},
	\end{equation}
	\begin{equation}
	\hat\rho_{ee'}^{n',(k+1)} = \frac{\sum_{t^{e'}_j} \sum_{t^e_i : t^{e'}_j<t^e_i} p^{(k)}_{i,j,n'}}{\sum_{t^{e'}_j} A^{n'}_{ee'}(j)} ,
	\end{equation}
	and setting $\hat\beta^{(k+1)}_{ee'}$ solution of Equation \eqref{eq:beta_update_equation}.
	\item Set $\hat\alpha^{n',(k+1)}_{ee'}=\hat\rho^{n',(k+1)}_{ee'}\hat\beta^{n',(k+1)}_{ee'}$.
	\item Set $k\leftarrow k+1$ and repeat from 2. until some convergence criterium is reached.
\end{enumerate}

\section{Complementary material for numerical illustrations}
\label{appendix:ComplNum}

This section provides some details on the numerical experiments reported in Section \ref{subsec:numerical}. For the first experiment, true values are given in Table \ref{table:EstimationExamples}. For the second and third experiments, true values of the parameters are
\begin{equation}
	\nu_1 = 0.5, 
	\, \nu_2 = 0.25, 
	\, \theta^1_1=0.25, 
	\, \theta^1_2=-0.25,
	\, \theta^2_1=-0.25,
	\, \theta^2_2=0.25,
\end{equation}	
and
\begin{equation}
	\alpha = \left(\begin{matrix}
	(5.0, 2.0, 0.1) & (5.0, 2.0, 0.1)
	\\ (10.0, 2.0, 0.2) & (10.0, 2.0, 0.2)
	\end{matrix}\right),
	\, \beta = \left(\begin{matrix}
	(50.0, 10.0, 1.0) & (100.0, 20.0, 2.0)
	\\ (200.0, 40.0, 4.0) & (100.0, 20.0, 2.0)
	\end{matrix}\right),
\end{equation}
where tuple notation in matrices $\alpha$ and $\beta$ denotes the multiple exponential parameters ($d_n=3$).

\begin{figure}
\centering
\includegraphics[width=0.5\textwidth]{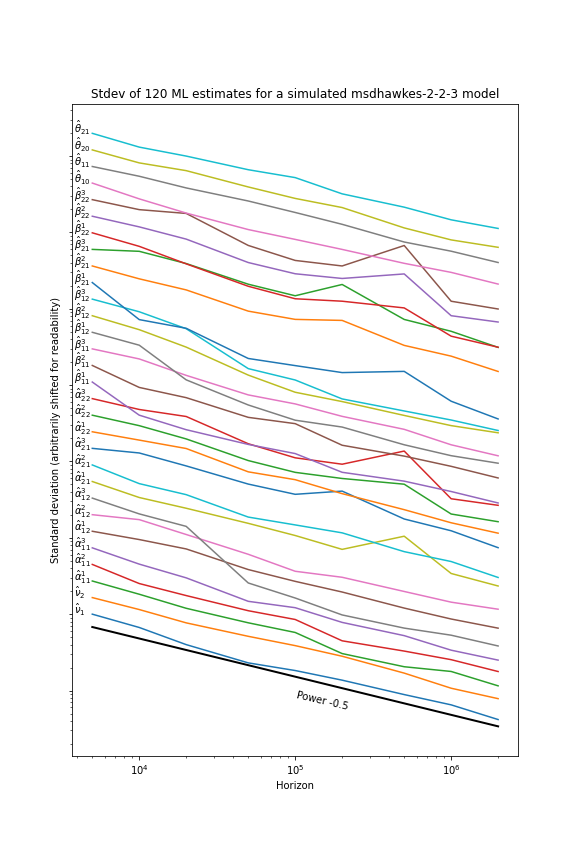}
\caption{Standard deviation of estimates of a msdHawkes model with $d_e=2$, $d_x=2$ and $d_n=3$. For readability, curves have been arbitrarily shifted along the $y$-axis.}
\label{fig:StdDeviationConvergence}
\end{figure}

As a complement to the second experiment, Figure \ref{fig:StdDeviationConvergence} plots the evolution of the standard deviation of the 30 estimates of a msdHawkes model with $d_e=2$, $d_x=2$ and $d_n=3$ as a function of the horizon $T$. This corresponds to the experiment described at Figure \ref{fig:EstimationExample} in the main text. Expected convergence in $\mathcal O(T^{-1/2})$ is retrieved.

Finally, for the fourth experiment, we use $d_e=2$, $d_x=2$, and kernels are of the form $\phi_{ee'}(t) = \alpha_{ee'}\left(1+t/\tau_{ee'}\right)^{-(1+\beta_{ee'})}$. The true values of the parameters of this fourth experiment are :
\begin{equation}
	\nu_1 = 0.5, 
	\, \nu_2 = 0.5, 
	\, \theta^1_1=0.25, 
	\, \theta^1_2=-0.5,
	\, \theta^2_1=-0.25,
	\, \theta^2_2=0.5,
\end{equation}	
and
\begin{equation}
	\alpha = \left(\begin{matrix}
	0.5 & 0.25
	\\ 0.25 & 0.5
	\end{matrix}\right),
	\, \beta = \left(\begin{matrix}
	1.0 & 2.0
	\\ 2.0 & 1.0
	\end{matrix}\right),
	\, \tau = \left(\begin{matrix}
	1.0 & 1.0
	\\ 1.0 & 1.0
	\end{matrix}\right).
\end{equation}

\section{Goodness-of-fit tests for msdHawkes processes}
\label{appendix:Residuals}
In this section, we provide computationally efficient forms for the residuals $r^e_i$ used to assess the goodness-of-fit of the msdHawkes models described in Section \ref{sec:msdHawkes}. We use the notations from there and from the previous appendix sections. We have for $i \in \intervalleEntier{n_e -1}$: 
\begin{equation}
r_i^e = \sum_{\substack{j \in \lbrace 1,...,N-1 \rbrace \\ t_i^{e'} \leqslant \tau_j < t_{i+1}^{e'}}}e^{\left< x_j,\hat{\theta^e} \right>}\int_{\tau_j}^{\tau_{j+1}}\hat{\lambda}^{H,e}(s)\diff s.
\end{equation}
With some computations, we show that for any $i \in \intervalleEntier{n_e -1}$ and $j \in \intervalleEntier{N-1}$ such that $t_i^{e'} \leqslant \tau_j < t_{i+1}^{e'}$, the integral term satisfies:
\begin{align}
    \int_{\tau_j}^{\tau_{j+1}}\hat{\lambda}^{H,e}(s)\diff s 
    & = \hat{\nu}_e\left(\tau_{j+1} - \tau_j \right) + \sum_{e'=1}^{d_e} \sum_{n=1}^{d_n} \frac{1}{\hat{\beta}_{ee'}^n}\bigg(\int_{]0,\tau_{j}]}\hat{\alpha}_{ee'}^n e^{-\hat{\beta}_{ee'}^n (\tau_{j}-u)}\diff N_u^{e'}
    \nonumber \\
    & \qquad\qquad\qquad\qquad\qquad\qquad\qquad - \int_{]0,\tau_{j+1}[}\hat{\alpha}_{ee'}^n e^{-\hat{\beta}_{ee'}^n(\tau_{j+1}-u)}\diff N_u^{e'}\bigg)
	\\
    & = \hat{\nu}_e\left(\tau_{j+1} - \tau_j \right) + \sum_{e'=1}^{d_e} \left(\hat{\mu}^{ee'}(\tau_j +)-\hat{\mu}^{ee'}(\tau_{j+1} -)\right),
\end{align}
where $\hat{\mu}^{ee'}$ is the intensity of a standard one-dimensional Hawkes process with kernel:
\begin{equation}
    \sum_{n=1}^{d_n} \frac{\alpha_{ee'}^n}{\beta_{ee'}^n} \exp\left(-\beta_{ee'}^n t \right),
\end{equation}
(any baseline intensity will do) and whose jump times are $\mathcal{T}^{e'}$.

\section{MLE and multiple events at the same timestamp}
\label{appendix:miniExample}

Having more than one event at occurring at the exact same time has probability 0 to happen in a simple intensity-driven point process. We show with a little example in one dimension why a sample with multiple events at the same timestamp is problematic for the maximum-likelihood estimation. Let $\{t_1, t_2\}$ a realization on $[0, T]$ of a standard Hawkes process with baseline intensity $\nu$ and kernel $\phi(t) = \alpha e^{-\beta t}$ (where $(\nu, \alpha, \beta)$ are the positive parameters to estimate by maximising the likelihood). Let's note $\epsilon = t_2 - t_1$, which will decrease to 0. The intensity of the process is:
\begin{equation}
    \forall t \geqslant 0,\quad \lambda_{\epsilon}(t) = \left\{\begin{array}{lll}
        \nu & \text{if} & t \in [0, t_1], \\
        \nu + \alpha e^{-\beta (t-t_1)} & \text{if} & t \in (t_1, t_1+\epsilon], \\
        \nu + \alpha e^{-\beta (t-t_1)} + \alpha e^{-\beta (t-t_1 - \epsilon)} & \text{if} & t \in (t_1 + \epsilon, T].
    \end{array} \right.
\end{equation}
By Equation \eqref{eq:pointProcessLogLik}, the log-likelihood of this realization is:
\begin{equation}
\begin{split}
    L_{\epsilon}(\nu, \alpha, \beta) = &\log(\nu) + \log\left(\nu + \alpha e^{-\beta\epsilon}\right) - \nu T \\
    &  - \frac{\alpha}{\beta} \left( 2 - e^{-\beta (T - t_1)} - e^{-\beta (T - t_1 - \epsilon)} \right).
\end{split}
\end{equation}
If we set $\epsilon = 0$, the log-likelihood becomes:
\begin{equation}
    L_{0}(\nu, \alpha, \beta) = \log(\nu) + \log\left(\nu + \alpha \right) - \nu T - 2 \frac{\alpha}{\beta} \left( 1 - e^{-\beta (T - t_1)} \right).
\end{equation}
By taking $\alpha = n$, $\beta = 2n$, $\nu = 1$ we have:
\begin{equation}
    \forall n \in \mathbb{N}^*, \sup_{(\nu, \alpha, \beta) \in \left(\mathbb{R}_+^*\right)^3}L_{0}(\nu, \alpha, \beta) \geqslant \log(1 + n) - T - 1 \xrightarrow[n \to +\infty]{} +\infty,
\end{equation}
thus the likelihood is not upper-bounded, so it does not have a maximum.
\end{document}